\documentclass[twocolumn]{aastex62}

\usepackage{txfonts}
\usepackage{graphicx,bm}
\usepackage{color,ulem}
\usepackage{todonotes}

\begin{document}

\title{Nucleosynthesis of an $11.8\,M_\odot$ Supernova with 3D Simulation of the Inner Ejecta:
Overall Yields and Implications for Short-Lived Radionuclides in the Early Solar System}
\shortauthors{Sieverding et al.}

\shorttitle{Nucleosynthesis of an $11.8\,M_\odot$ Supernova}

\author{A. Sieverding}
\affiliation{School of Physics and Astronomy,
      University of Minnesota, Minneapolis, MN 55455, USA} 

\author{B.~M\"uller}
\affiliation{School of Physics and
  Astronomy, Monash University, Victoria 3800, Australia} 

\author{Y.-Z. Qian}
\affiliation{School of Physics and Astronomy,
      University of Minnesota, Minneapolis, MN 55455, USA} 

\begin{abstract}
    Based on a 3D supernova simulation of an $11.8\,M_\odot$ progenitor model with initial
    solar composition, we study the nucleosynthesis using tracers covering the 
    innermost $0.1\,M_\odot$ of the ejecta. These ejecta are mostly proton-rich and contribute
	significant amounts of $^{45}$Sc and $^{64}$Zn. The production of heavier isotopes is
	sensitive to the electron fraction and hence the neutrino emission from the proto-neutron star.
	The yields of these isotopes are rather uncertain due to the approximate neutrino transport 
	used in the simulation. In order to obtain the total yields for the whole supernova, 
	we combine the results from the tracers with those for the outer layers from a suitable 1D
	model. Using the yields of short-lived radionuclides (SLRs), we explore the possibility that
	an $11.8\,M_\odot$ supernova might have triggered the formation of the solar system and 
	provided some of the SLRs measured in meteorites. In particular, we discuss two new scenarios 
	that can account for at least the data on $^{41}$Ca, $^{53}$Mn, and $^{60}$Fe without 
	exceeding those on the other SLRs.
\end{abstract}


\section{Introduction}
\label{sec:intro}
Core-collapse supernovae (CCSNe), which mark the end of the lives of massive
stars, are important sites for nucleosynthesis by producing 
most of the elements up to and including the Fe group
\citep[e.g.,][]{Nomoto.Kobyashi.ea:2013,Thielemann.Nomoto.ea:1996}.  
Parameterized models based on pistons or thermal energy injections
in spherical symmetry have been widely used to guide our understanding of 
CCSN contributions to chemical enrichment of the universe
\citep[e.g.,][]{Woosley.Heger.ea:2002,Heger.Woosley:2010}.
Being computationally efficient, such models can be calibrated to well-known 
observables and then be employed to explore fine grids of initial mass, metallicity, 
and rotation \citep[e.g.,][]{Limongi.Chieffi:2018}.
They, however, have proven inadequate for predicting the conditions of 
the inner CCSN ejecta and hence the associated nucleosynthesis
\citep{Wanajo.Mueller.ea:2018,Harris.Hix.ea:2017,Lentz.Bruenn.ea:2015,Young.Fryer:2007}.
With an improved understanding of the CCSN mechanism
\citep[e.g.,][]{Lentz.Bruenn.ea:2015,Janka.Melson.ea:2016,Mueller:2016,OConnor.Couch:2018,Burrows.Radice.ea:2020}, advanced
parameterizations for explosions in spherical symmetry have been developed
\citep{Ebinger.Curtis.ea:2019,Ertl.Janka.ea:2016}. Yet it remains unclear
whether such parameterizations can accurately predict the conditions of the
ejecta directly affected by the dynamics of the neutrino-driven explosion. 
Fully self-consistent multidimensional simulations are important benchmarks for such models.

A number of 3D simulations have been performed recently 
\citep{Burrows.Radice.ea:2020,Glas.Just.ea:2018,Mueller.Tauris.ea:2019},
but detailed studies of the nucleosynthesis so far have been based mostly on 2D models
\citep{Eichler.Nakamura.ea:2018,Wanajo.Mueller.ea:2018,Harris.Hix.ea:2017}.
It has become clear, however, that there are systematic differences between the 
dynamics of 2D and 3D explosions \citep{Mueller:2015}, which might affect the nucleosynthesis.
Therefore, 3D simulations are required to address definitively the nucleosynthesis of
the inner ejecta that are directly affected by the explosion mechanism, especially
the production of the Fe-group and heavier elements.
Reliable nucleosynthesis predictions require running 3D simulations until several seconds 
after core bounce, which, however, is not done in most cases due to the extremely high 
computational cost.

In this study, we calculate the nucleosynthesis of an $11.8\,M_\odot$ CCSN based on a 
3D simulation with the CoCoNuT-FMT code \citep{Mueller.Tauris.ea:2019,Mueller.Janka.ea:2015}.
The simulation was run until $\approx 1.2\,\mathrm{s}$ after core bounce, 
long enough to cover most of the explosive nucleosynthesis for the whole Si shell 
and part of the O/Ne layer. Using 2D models, \citet{Wanajo.Mueller.ea:2018} already pointed 
out differences in the yield pattern from such inner ejecta as compared to commonly adopted 
parameterizations in spherical symmetry. For progenitors with masses similar to the one 
studied here, they found larger production factors of $^{45}$Sc and $^{64}$Zn because
neutrino interaction results in a larger variability of the electron fraction $Y_\mathrm{e}$ in 2D.
We confirm this finding for the first time with a 3D model. 
Regarding the production of heavier isotopes, however, we find that to reach firm conclusions,
our simulation must be improved by incorporating more accurate neutrino transport to determine 
self-consistently the $Y_\mathrm{e}$ of the neutrino-heated ejecta.

To obtain the total yields of the $11.8\,M_\odot$ CCSN, we combine the results for the inner 
ejecta from the 3D model with those for the outer layers from a suitable explosion model in 
spherical symmetry. Using the yields of short-lived radionuclides 
(SLRs, with lifetimes up to several $10^7$ yr),
we explore the possibility that such a CCSN might have triggered the formation of the solar 
system and provided some of the SLRs measured in meteorites.
An $11.8\,M_\odot$ progenitor model with an initial solar composition 
(corresponding to a total mass fraction of 0.014 for metals, \citealt{Asplund.Grevesse.ea:2009})
is of special interest. \citet{Banerjee.Qian.ea:2016} 
showed that if a reasonably small amount of the inner material is allowed to fall back onto 
the proto-neutron star (PNS), the yields from a parameterized 1D explosion of this model 
are consistent with the meteoritic data and can account for the abundances of the SLRs
$^{10}$Be, $^{41}$Ca, and $^{107}$Pd in the early solar system (ESS). 
In contrast, typical CCSNe from more massive stars tend to grossly overproduce the SLRs 
$^{53}$Mn and $^{60}$Fe and cause large shifts in stable isotopes that are not observed.
At the end of our 3D CCSN simulation, we find that the yields of $^{53}$Mn and $^{60}$Fe of 
the $11.8\,M_\odot$ model are comparable to those obtained by \citet{Banerjee.Qian.ea:2016}
without fallback. Because neither our 3D simulation nor our 1D model shows any indication of
fallback, we discuss two new scenarios that can account for at least the ESS data on $^{41}$Ca, 
$^{53}$Mn, and $^{60}$Fe without exceeding those on the other SLRs.
\citet{Banerjee.Qian.ea:2016} also showed that $^{10}$Be is commonly produced by CCSNe.
We study the production of this SLR by the $11.8\,M_\odot$ model
in detail and identify several sources of uncertainty.

The plan of this paper is as follows. In \S\ref{sec:method} we discuss our method of extracting 
Lagrangian tracers from the 3D simulation and the setup for the nucleosynthesis calculations. 
In \S\ref{sec:nucleo_inner} we give an overview of the nucleosynthesis of the inner ejecta
calculated with the tracers. In \S\ref{sec:nucleo_outer} we present the yields for the entire CCSN
combining the results for the inner ejecta and those for the outer ejecta calculated with an
appropriate 1D model. We also discuss the production of the SLR $^{10}$Be in detail.
In \S\ref{sec:SLRs} we focus on the SLR yields and discuss their implications for the ESS.
In \S\ref{sec:conclusions} we summarize our results and give conclusions.

\section{Method}
\label{sec:method}
The $11.8\,M_\odot$ progenitor model with an initial solar composition was evolved
with the 1D hydrodynamics code KEPLER \citep{Weaver.Zimmerman.Woosley:1978} by \citet{Banerjee.Qian.ea:2016} 
to explore candidates for a possible CCSN trigger for the formation of the solar system. \citet{Mueller.Tauris.ea:2019} 
carried out a 3D CCSN simulation based on this model. We refer readers to that work for details 
of the simulation. We describe in this section how we extract the conditions of the inner ejecta
from the simulation and how we calculate the nucleosynthesis of the CCSN.

\subsection{Tracer Extraction}
\label{sec:extraction}
\begin{figure} \centering
	\includegraphics[width=\linewidth]{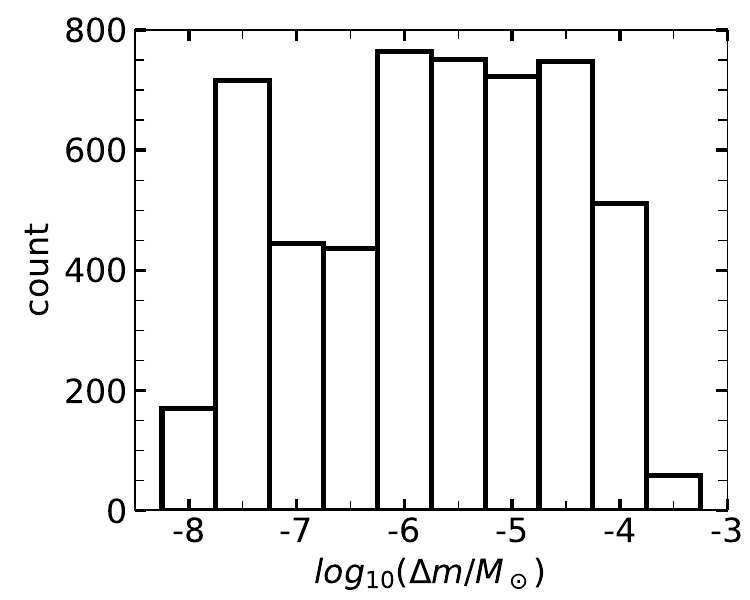}
	\caption{Histogram for the number of tracers as a function of their masses.}
\label{fig:mass_resolution} 
\end{figure}

At the end of the 3D CCSN simulation, we place tracers on a spatial grid that is equally spaced 
in $\log_{10}r$, $\cos\theta$, and $\phi$, where $r$ is the radius ranging from 100 to 15,000~km, 
$\theta$ is the polar angle, and $\phi$ is the azimuthal angle. We assign a mass to each tracer by
integrating the mass density over the corresponding cell.
Consequently, the masses of the tracers are not equal but reflect the density distribution
at the end of the simulation. Further, due to the equal spacing in $\log_{10}r$, the mass
resolution ranges from $\Delta m\sim1.3 \times 10^{-8}\,M_\odot$ for the innermost tracers
to $\Delta m\sim2.2\times 10^{-4}\,M_\odot$ for the outermost ones.
The distribution of tracer masses is shown in Figure~\ref{fig:mass_resolution}.  

Starting from the final position at the end of the simulation, we reconstruct
the trajectory of each tracer backward in time based on snapshots of the simulation 
in intervals of 2~ms until 100~ms after core bounce. A similar approach was used by \citet{Wanajo.Mueller.ea:2018} 
for 2D simulations. This post-processing is less accurate than tracking the tracers 
during the simulation on the fly, which typically takes much shorter time steps.
On the other hand, this procedure takes advantage of the separation between the 
ejecta and the material ending up in the PNS, 
so the tracers can be distributed more efficiently.
In addition, while numerical errors accumulated during the integration
affect the accuracy of a trajectory at early times, the affected part of the trajectory
typically corresponds to high temperatures of $T\gtrsim 6.5$~GK for which nuclear 
statistical equilibrium (NSE) holds. Because nucleosynthesis mostly depends on the 
evolution after the freeze-out from NSE, it is not much affected by such inaccuracy.

\begin{figure}
 \includegraphics[width=\linewidth]{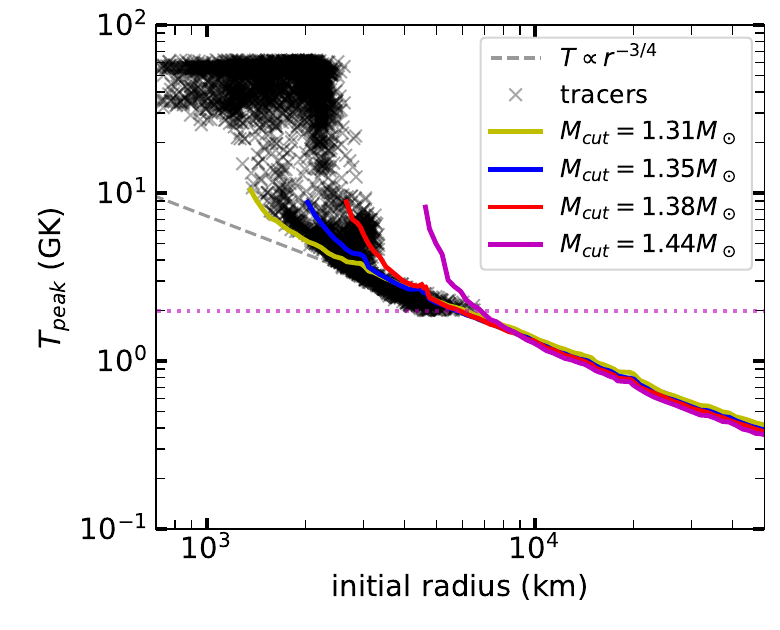}
 \caption{\label{fig:tpeak_comparison} Peak temperature as a function of initial radius. 
 The results for different 1D models are shown as curves. The dotted horizontal line indicates 
 the minimum peak temperature of $2\,\mathrm{GK}$ that we require of the selected tracers.
 See \S\ref{sec:nucleo_general} for detailed explanation.}
\end{figure}

Due to ambiguities about the final fate, uncertainties in extrapolation, and numerical 
integration errors, we cannot use all the extracted tracer histories, but select
the tracers based on the following three criteria.
\begin{enumerate}
    \item We only select tracers that show a positive radial velocity
averaged over the last 10~ms of the simulation.\footnote{
Material with a negative total energy at the end of the simulation still has a 
chance of being ejected as the energy is raised by pressure forces and turbulent 
viscous drag of the expanding bubbles. Therefore, we do not use the total energy
as a criterion to determine the final fate of a tracer.} 
Tracers with a negative radial velocity are assumed to become part of 
the PNS eventually.
\item We only select tracers that reach a temperature above $2\,\mathrm{GK}$.  
This criterion excludes those tracers that either have 
not been shocked or have too few post-shock data points for reliable 
extrapolation. They are found at the outer edge of the simulation domain. 
We assume that the corresponding material can be reasonably well described by 
a model in spherical symmetry as indicated in Figure \ref{fig:tpeak_comparison} 
and are be explained in \S\ref{sec:nucleo_general}.
\item We reject those tracers that have a change of entropy exceeding 50\% between 
successive time steps during the last 400~ms of the evolution, which most likely 
results from integration errors. We still expect this material to be part of 
the ejecta and redistribute its total mass among the selected tracers, which
increases the masses of the latter by a uniform factor of 1.11.
\end{enumerate}

In the end, we have selected 5790 tracers representing $0.1\,M_\odot$ of the innermost ejecta.
We have checked that the distributions of entropy and $Y_\mathrm{e}$ for these ejecta are not noticeably 
affected by the rejection and mass redistribution in our selection of the tracers.
We refer readers to \citet{Harris.Hix.ea:2017}
for a detailed discussion of the issues that arise from the initial placement and
final selection of tracers. 

\subsection{Tracer Extrapolation} 
The simulation is run long enough to determine the PNS properties and important 
parameters of the explosion. Nucleosynthesis, however, would occur in the 
expanding material significantly beyond the time $t=t_\mathrm{e}$ when the simulation ends.
In order to calculate the yields, we need to extrapolate the evolution of the radius, 
temperature, and density of each tracer. We assume that at $t>t_\mathrm{e}$,
each tracer moves with a fixed velocity $v_r(t_\mathrm{e})$, which is determined from the 
finite-difference approximation for its $dr/dt$ over the last 5~ms of the simulation.
Under this assumption, the radius of a tracer expands linearly as
$r(t)=r(t_\mathrm{e})+v_r(t_\mathrm{e})(t-t_\mathrm{e})=r(t_\mathrm{e})[1+(t-t_\mathrm{e})/\tau]$, where $\tau=r(t_\mathrm{e})/v_r(t_\mathrm{e})$.
We further assume that its density and temperature evolve as
(e.g., \citealt{Ning.Qian.ea:2007}) 
\begin{equation}
\label{eq:power_law} \rho(t)=\rho(t_\mathrm{e}) \left[ 1+(t-t_\mathrm{e})/\tau  \right]^{-2} ,
\end{equation} 
and 
\begin{equation} \label{eq:power_law_T}
	T(t)=T(t_\mathrm{e}) \left[ 1+(t-t_\mathrm{e})/\tau  \right]^{-2/3} , 
\end{equation} 
respectively. The above extrapolation is far from unique and other
authors have uses different functional forms. We have also performed calculations 
with an exponential expansion for $r(t)$ and found no major difference in the 
nucleosynthesis from the above linear expansion. 

\subsection{Neutrino Luminosities and Spectra}
Neutrinos are critical to the determination of the $Y_\mathrm{e}$ of the inner ejecta and 
play direct roles in the production of the SLR $^{10}$Be and other nuclei.
Neutrino luminosities and average energies are available from the simulation only
for the first $\approx 1.2\,\mathrm{s}$, over which time the energy emitted in neutrinos
is $4.2\,\times 10^{52}$ erg. We extrapolate the luminosity for each neutrino species
as $L_{\nu_i}(t)=L_{\nu_i}(t_\mathrm{e})\exp[-(t-t_\mathrm{e})/\tau_\nu]$ beyond the end of the
simulation. The decay timescale $\tau_\nu$ is determined by setting the total
energy emitted in neutrinos to be the gravitational binding energy of the PNS:
\begin{equation} 
\sum_{\nu_i}\int_0^\infty L_{\nu_i}(t) dt=(M_{\mathrm{PNS}}-M_G)c^2,
\end{equation}
where $M_{\mathrm{PNS}}$ and $M_G$ are the baryonic and gravitational mass of
the PNS, respectively. Following \citet{Mueller:2015}, we take
\begin{equation}
M_G=\frac{\sqrt{1+(0.336 M_{\mathrm{PNS}}/M_\odot)}-1}{0.168} M_\odot,
\end{equation} 
which corresponds to a binding energy of $2.25 \times 10^{53}\,\mathrm{erg}$ for
$M_{\mathrm{PNS}}=1.35M_\odot$. The above procedure gives $\tau_\nu=2$~s.

In addition, we assume that the average neutrino energies 
$\langle \varepsilon_{\nu_i} \rangle$ linearly decrease
at $t>t_\mathrm{e}$, reaching zero at $10\,\mathrm{s}$ after core bounce. 
The neutrino spectra are important for the determination of $Y_\mathrm{e}$ 
and for the $\nu$ process. For the neutrino reactions on nucleons, 
the selected reactions listed in \citet{Sieverding.Langanke.ea:2019}, 
and all the reactions affecting the $^{10}$Be yield, we follow   
\citet{Keil.Raffelt.ea:2003} and  \citet{Tamborra.Mueller.ea:2012} 
and take the normalized neutrino spectra to be
\begin{equation}
  \label{eq:alpha_fit}
    n_{\nu_i} (E) \equiv \left(\frac{\alpha+1}{\langle
      \varepsilon_{\nu_i}\rangle}\right)^{\alpha+1}\frac{E^\alpha}{\Gamma(\alpha+1)}\exp\left(-\frac{(\alpha+1)
      E}{\langle \varepsilon_{\nu_i} \rangle}\right),
\end{equation}
where $\Gamma(\alpha+1)$ is the $\Gamma$ function, and $\alpha$ is the so-called ``pinching'' parameter.
We fix $\alpha$ throughout the calculations and use $\alpha=3.30$, 3.12, and 2.3 for $\nu_\mathrm{e}$, $\bar{\nu}_\mathrm{e}$, 
and $\nu_x$ ($x=\mu$, $\tau$; $\nu_x$ and $\bar\nu_x$ have the same emission properties), respectively.
For all the other neutrino-induced reactions, we assume Fermi-Dirac spectra with zero chemical potential
and the corresponding average energies $\langle\varepsilon_{\nu_i}\rangle$.

\subsection{Corrections of Neutrino Emission}
\label{sec:lums}
\begin{figure}[ht]
	\includegraphics[width=\linewidth]{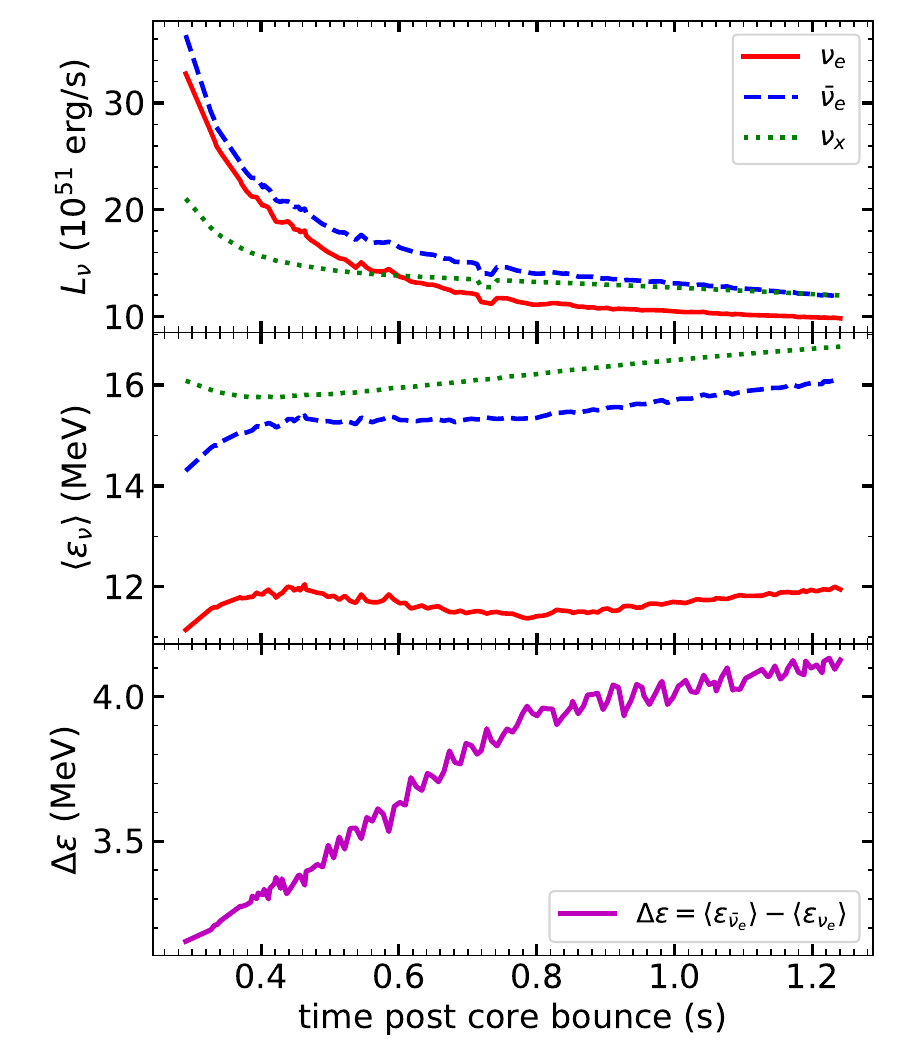}
	\caption{\label{fig:nu_signal} Time evolution of angle-averaged neutrino luminosities and
	average energies from the 3D simulation of the $11.8\,M_\odot$ CCSN.
	The bottom panel shows the difference in average energy between
	$\nu_\mathrm{e}$ and $\bar{\nu}_\mathrm{e}$, which is crucial for determining the $Y_\mathrm{e}$ of the inner ejecta
	and hence the associated nucleosynthesis.} 
\end{figure}

Supernova simulations with state-of-the-art neutrino transport predict that
$\nu_\mathrm{e}$ and $\bar{\nu}_\mathrm{e}$ have almost identical luminosities but somewhat different 
average energies with $\Delta\varepsilon=\langle \varepsilon_{\bar{\nu}_\mathrm{e}} \rangle - \langle 
\varepsilon_{\nu_\mathrm{e}} \rangle \approx 2\,\mathrm{MeV}$ during the first second after 
core bounce \citep{Mueller:2019}. Because the present 3D simulation used simplified 
neutrino transport \citep{Mueller.Janka.ea:2015}, it gives larger values of 
$\Delta\varepsilon\sim 3$--4~MeV (see Figure \ref{fig:nu_signal}).
As pointed out in \citet{Mueller.Tauris.ea:2019}, this caveat leads to inaccurate determination
of the $Y_\mathrm{e}$ of the inner ejecta by the simulation while the overall dynamics of the explosion 
is much less affected by the exact differences between the $\nu_\mathrm{e}$ and $\bar{\nu}_\mathrm{e}$ emission. 
In order to obtain reasonable results of nucleosynthesis,
we apply corrections to the neutrino emission properties from the simulation and recalculate the 
evolution of the $Y_\mathrm{e}$ for the selected tracers.

Modifications of the neutrino emission should be consistent with the original hydrodynamic
evolution of the tracers. For this purpose, we keep approximately the same 
neutrino heating rate, which is proportional to that of \citet{Qian.Woosley:1996}:
\begin{equation}
\label{eq:heating}
   Q=L_{\nu_\mathrm{e}} \frac{\langle \varepsilon_{\nu_\mathrm{e}}^3 \rangle}{\langle \varepsilon_{\nu_\mathrm{e}} \rangle}
    +L_{\bar{\nu}_\mathrm{e}} \frac{\langle \varepsilon_{\bar{\nu}_\mathrm{e}}^3 \rangle}{\langle \varepsilon_{\bar{\nu}_\mathrm{e}} \rangle},   
\end{equation}
with $\langle \varepsilon_{\nu_\mathrm{e}}^3 \rangle=1.81 \langle \varepsilon_{\nu_\mathrm{e}} \rangle^3$ and 
$\langle \varepsilon_{\bar{\nu}_\mathrm{e}}^3 \rangle=1.85 \langle \varepsilon_{\bar{\nu}_\mathrm{e}} \rangle^3$ 
being the third moments of the corresponding spectra. To obtain the modified $\nu_\mathrm{e}$ and $\bar\nu_\mathrm{e}$
emission properties, which are denoted as the primed quantities below, we specify
\begin{equation}
\langle \varepsilon_{\nu_\mathrm{e}} \rangle' = \bar{\varepsilon} - \frac{1}{2}\Delta \varepsilon,\quad
\langle \varepsilon_{\bar{\nu}_\mathrm{e}} \rangle' = \bar{\varepsilon} + \frac{1}{2}\Delta \varepsilon,
\end{equation}
where $\Delta \varepsilon=2$~MeV, $\bar{\varepsilon} = (L_{\nu_\mathrm{e}} +L_{\bar\nu_\mathrm{e}})/(N_{\nu_\mathrm{e}}+N_{\bar\nu_\mathrm{e}})$,
and for example, $N_{\nu_\mathrm{e}}=L_{\nu_\mathrm{e}}/\langle \varepsilon_{\nu_\mathrm{e}} \rangle$.
We keep the same pinching parameters $\alpha$ as specified above for the neutrino spectra.
Then the modified $\nu_\mathrm{e}$ and $\bar\nu_\mathrm{e}$ luminosities are obtained from
\begin{equation}
 \label{eq:modification}
 L_{\nu_\mathrm{e}}'=L_{\bar{\nu}_\mathrm{e}}'=Q \left[ 
 \frac{
  \langle \varepsilon_{\nu_\mathrm{e}}^3 \rangle'
      }{
      \langle \varepsilon_{\nu_\mathrm{e}} \rangle'
      } + 
      \frac{
  \langle \varepsilon_{\bar{\nu}_\mathrm{e}}^3 \rangle'
      }{
      \langle \varepsilon_{\bar{\nu}_\mathrm{e}} \rangle'
      }      
      \right]^{-1}.
\end{equation}
The above modifications slightly change the luminosities of $\nu_\mathrm{e}$ and $\bar\nu_\mathrm{e}$ 
but enforce the desired difference between their average energies while keeping
the neutrino heating rate consistent with the hydrodynamics of the simulation.

\begin{figure*}
    \centering
    \includegraphics[width=\linewidth]{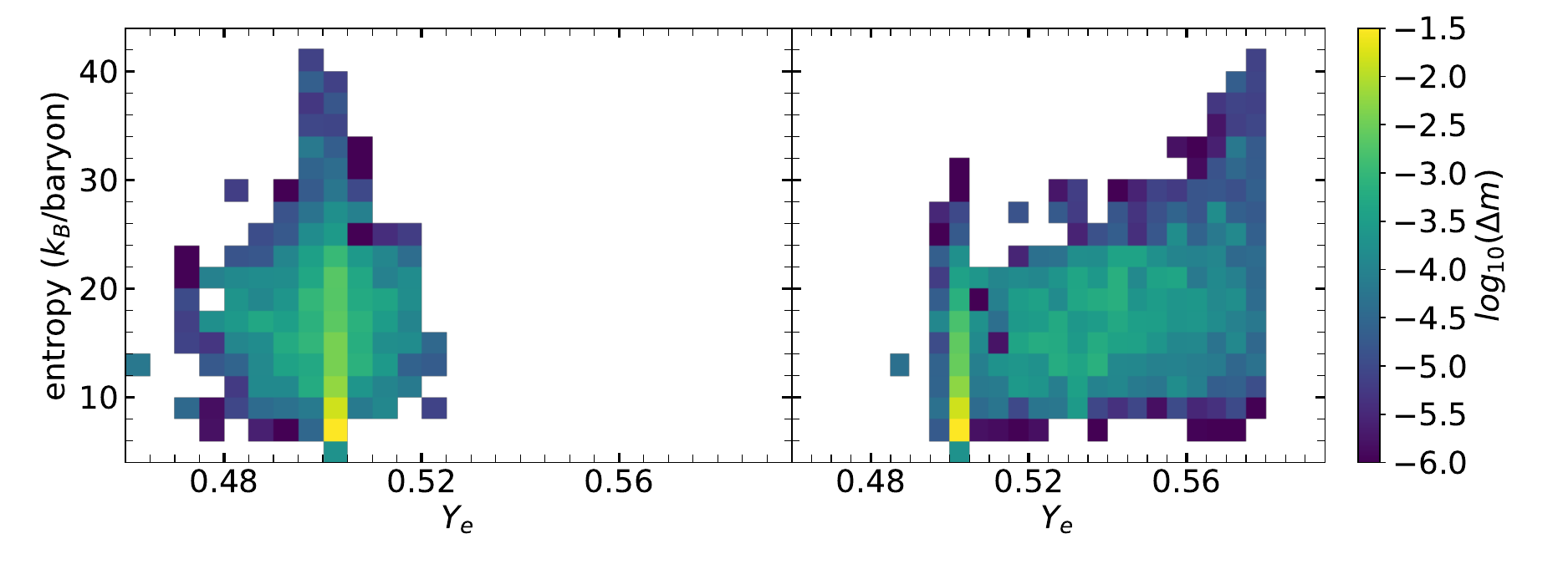}
    \caption{Distributions of tracer mass in $Y_\mathrm{e}$ and entropy assuming the original (left panel) 
    and modified (right panel) neutrino emission properties.
    The prominent presence of matter with $Y_\mathrm{e}\approx 0.5$ and low entropy can be
    traced to its being shocked at large radii.
   }
    \label{fig:ye-s-plane}
\end{figure*}

As discussed in \S\ref{sec:nucleo_general}, we follow the evolution of the $Y_\mathrm{e}$ for each tracer 
by including the pertinent weak reactions, especially $\nu_\mathrm{e}$ and $\bar\nu_\mathrm{e}$ absorption on free nucleons,
in our nucleosynthesis calculations. We also evaluate the entropy of each tracer using its temperature,
density, and composition.
Figure \ref{fig:ye-s-plane} compares the distributions of tracer mass in $Y_\mathrm{e}$ and entropy 
for the original (left panel) and modified (right panel) neutrino luminosities and average energies. 
The values of $Y_\mathrm{e}$ are extracted when the temperature drops to $6.5\,\mathrm{GK}$ and those of entropy 
(in units of Boltzmann constant $k_B$ per baryon) are
taken when the temperature drops to $2\,\mathrm{GK}$ for the last time.
The reduced $\Delta \varepsilon$ extends the $Y_\mathrm{e}$ range to higher values.
This extension is in agreement with other CCSN 
simulations for similar progenitors, which have found mostly proton-rich ejecta 
\citep{Mueller.Janka.ea:2012,Wanajo.Mueller.ea:2018,Vartanyan.Burrow.ea:2019}. In contrast,
the modified neutrino emission causes little change in the range of entropy, which is mostly determined by
the temperature and density. These two quantities are taken to be unaffected by the modifications
because we keep the same neutrino heating rate. With the modified neutrino emission, however,
higher values of $Y_\mathrm{e}$ tend to be associated with higher values of entropy.
This trend can be understood because both the $Y_\mathrm{e}$ and entropy are mainly set by $\nu_\mathrm{e}$ and $\bar\nu_\mathrm{e}$ 
absorption on free nucleons. Higher entropy corresponds to more occurrences of these reactions, which tend
to raise the $Y_\mathrm{e}$ more significantly with the reduced $\Delta \varepsilon$.

\subsection{Nucleosynthesis Calculations}
\label{sec:nucleo_general}

In general, as the CCSN shock propagates further into the outer layers, the dynamical evolution 
of the shocked ejecta is expected to approach that in an equivalent 1D explosion model. 
Accordingly, we calculate the nucleosynthesis of the inner ejecta using tracers for
as much mass as allowed by the 3D simulation data and treat the outer layers using a 1D model 
consistent with the 3D simulation. As explained below, this approach is motivated by a comparison
of the peak temperature of the shocked ejecta between the 3D simulation and the 1D model.

Figure \ref{fig:tpeak_comparison} shows the peak temperature of the tracers 
as a function of their initial radii at $100\,\mathrm{ms}$ after core bounce.
There is a large spread in this temperature for tracers with small initial radii.\footnote{
Numerical integration errors arise when a tracer remains almost stationary at small radii for 
an extended period of time. Consequently, the initial position is rather uncertain 
for those tracers that move near or into the PNS during their evolution.} The spread,
however, is reduced drastically for tracers with initial radii of $r_0>3000\,\mathrm{km}$.
This result arises because the shock starts to evolve almost spherically once it breaks out 
of the high-density Fe core. As indicated by the dashed line in Figure~\ref{fig:tpeak_comparison},
for $r_0 > 3000 \,\mathrm{km}$, the peak temperature follows $T_{\mathrm{peak}}\propto r_0^{-3/4}$,
which corresponds to a spherical shock dominated by radiation.
For comparison, Figure \ref{fig:tpeak_comparison} also shows the peak temperature as a function 
of the initial radius for piston-driven spherical explosion of the same progenitor model 
with a mass cut $M_{\mathrm{cut}}=1.31$, 1.35, 1.38, and $1.44\,M_\odot$ and with an
explosion energy of $0.2\times10^{51}$, $0.15\times10^{51}$, $0.13\times10^{51}$,
and $0.1\times10^{51}$ erg, respectively. The results converge for $T_{\mathrm{peak}}\lesssim2$~GK,
which indicates that the conditions, and hence the nucleosynthesis, of the ejecta with
$T_{\mathrm{peak}} \lesssim 2\,\mathrm{GK}$ are practically the same for these 1D models.
On the other hand, the models with $M_{\mathrm{cut}}=1.31$ and $1.38\,M_\odot$ bracket
the peak temperature of the tracers up to $T_{\mathrm{peak}} \sim 9\,\mathrm{GK}$ while the one with
$M_{\mathrm{cut}}=1.35\,M_\odot$ gives the best description of the mean trend. The latter model
closely matches the PNS mass $M_{\rm PNS}=1.35\,M_\odot$ and the explosion energy of $0.2\times10^{51}$ erg 
found in the 3D simulation.
Based on the above discussion, we use tracers to calculate the nucleosynthesis of the inner 
ejecta with $T_{\mathrm{peak}}\geq 2$~GK and adopt the 1D model with 
$M_{\mathrm{cut}}=M_{\rm PNS}=1.35\,M_\odot$ to treat the outer ejecta.
As the total mass of the selected tracers, or the inner ejecta, is $M_{\rm in}=0.1\,M_\odot$, the transition between 
the inner and outer ejecta is at an enclosed mass of $M_{\rm tr}=M_{\rm PNS}+M_{\rm in}=1.45\,M_\odot$.

For both the inner and outer ejecta, the associated nucleosynthesis is calculated with 
the reaction network used in \citet{Sieverding.Martinez.ea:2018}, which includes 5300 isotopes. 
Thermonuclear reaction rates are taken from the REACLIB v2.2 library \citep{Cyburt.Amthor.ea:2010}.
When available, $\beta$-decay rates are taken from the NUBASE compilation of experimentally determined 
values \citep{Audi:2017}. Otherwise, theoretical $\beta$-decay rates from \citet{Moeller.Pfeiffer.ea:2003} 
are used. As in \citet{Sieverding.Langanke.ea:2019}, our reaction network includes a large number of 
neutrino-nucleus reactions, for which the rates are calculated with the appropriate neutrino spectra.
When NSE is achieved, the composition is fully determined by the temperature, density, and $Y_\mathrm{e}$. 
We assume NSE for temperatures above $6.5\,\mathrm{GK}$ and use the reaction network to follow the 
evolution of the composition at lower temperatures. 

Regardless of whether NSE is achieved, the evolution of $Y_\mathrm{e}$ is always followed with 
the relevant weak interactions, including neutrino reactions, e$^\pm$ capture, and $\beta$-decay.
The rates of e$^\pm$ capture on nuclei are taken from \citet{Langanke.Martinez:2005}. 
The most important weak reactions are
\begin{equation}
	\nu_\mathrm{e} + \mathrm{n} \rightleftharpoons \mathrm{e}^- + \mathrm{p},\quad
	\bar\nu_\mathrm{e} + \mathrm{p} \rightleftharpoons \mathrm{e}^+ + \mathrm{n}.
    \label{eq:weak_reactions}
\end{equation}
We calculate the rates for the above reactions following \citet{Bruenn:1985} and
using the axial-vector coupling constant $g_a=1.2783$ \citep{Brown.Dees.ea:2018}.
Pauli blocking of e$^\pm$ in the final state is taken into account and
detailed balance is enforced for the matrix elements of the forward and reverse
reactions. We also include the weak magnetism and nucleon recoil corrections \citep{Horowitz:2002}.
The $Y_\mathrm{e}$ is very sensitive to the competition between the forward and reverse reactions,
and that between the $\nu_\mathrm{e}$ and $\bar\nu_\mathrm{e}$ reactions in Equation~(\ref{eq:weak_reactions}). 
Relatively small changes in the relevant rates, such as the weak magnetism and nucleon recoil 
corrections, can produce significant changes in $Y_\mathrm{e}$ that affect nucleosynthesis.

\section{Nucleosynthesis of the Inner Ejecta}
\label{sec:nucleo_inner}

We discuss the nucleosynthesis of the inner ejecta in this section.
The calculation is based on the tracers for the 3D simulation as described in \S\ref{sec:method}, 
and covers a total of $0.1\,M_\odot$ of the inner ejecta. Due to the limited time,
the simulation does not cover matter ejected in the neutrino-driven winds\footnote{
These winds represent the material ejected from the vicinity of the PNS by
neutrino heating. Their nucleosynthesis depends on the characteristics of neutrino
emission mainly through the entropy, the expansion timescale, and especially the electron
fraction determined by neutrino heating. In contrast, the $\nu$ process included in
our study mainly takes place in the outer ejecta and depends on the characteristics 
of neutrino emission directly through their impact on the rates of neutrino-nucleus reactions.}
during the long-term cooling of the PNS. We refer readers to the extensive literature for 
the nucleosynthesis of the neutrino-driven winds 
\citep[e.g.,][]{Woosley.Hoffman.ea:1992,Qian.Woosley:1996,Hoffman.Woosley.Qian:1997,Roberts.Woosley.ea:2010,Bliss.Arcones.ea:2018}. 

\subsection{Conditions of the Inner Ejecta}
\label{sec:nucleo_conditions}

In general, tracers reaching temperatures of $6.5\,\mathrm{GK}$ or above evolve through 
a phase of NSE and their nucleosynthesis is sensitive to the conditions, especially the 
$Y_\mathrm{e}$, at the freeze-out from NSE. In contrast, the nucleosynthesis of those tracers 
never achieving NSE is largely determined by their peak temperatures and is also sensitive 
to the initial composition.

\begin{figure}[ht]
 \includegraphics[width=\linewidth]{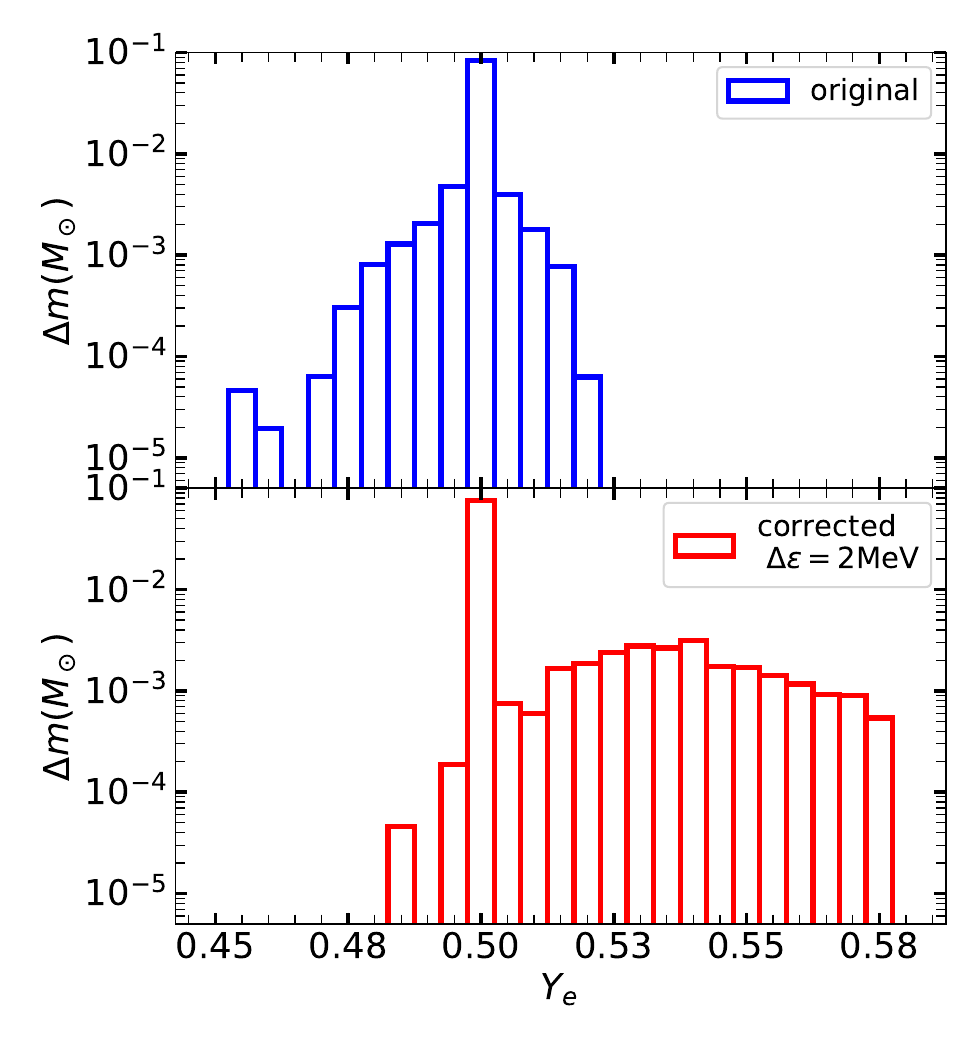}
 \caption{Distributions of tracer mass over the freeze-out values of $Y_\mathrm{e}$
 assuming the original (top panel) and modified (bottom panel) neutrino emission properties.
 \label{fig:ye-histo}}
\end{figure}

For tracers evolving through the NSE phase, their freeze-out values of $Y_\mathrm{e}$ 
are mostly set by the competition between 
the $\nu_\mathrm{e}$ and $\bar{\nu}_\mathrm{e}$ reactions with free nucleons in Equation~(\ref{eq:weak_reactions})
(e$^\pm$ captures on free nucleons are important only at relatively high densities). 
As soon as nuclei form, the efficiency of the $\nu_\mathrm{e}$ and $\bar{\nu}_\mathrm{e}$ reactions is suppressed
and the $Y_\mathrm{e}$ effectively stops changing. Figure \ref{fig:ye-histo} shows that with the original 
neutrino luminosities and spectra from the simulation, there is a relatively narrow distribution 
for the freeze-out values of $Y_\mathrm{e}$ centered around $Y_\mathrm{e}\approx0.5$ with tails extending down to 
neutron-rich conditions of $Y_\mathrm{e}\approx 0.45$ and up to proton-rich conditions of $Y_\mathrm{e}\approx 0.52$. 
Figure \ref{fig:ye-s-plane} also shows that the extreme values of $Y_\mathrm{e}$ mostly correspond to low values 
of $\lesssim 20\,k_\mathrm{B}/\mathrm{baryon}$ for the entropy. For tracers with $Y_\mathrm{e}\approx 0.5$,
their values of entropy have a wider range and can reach up to $41 k_\mathrm{B}/\mathrm{baryon}$.

With the corrections of neutrino emission described in \S\ref{sec:lums}, 
the distribution of $Y_\mathrm{e}$ is spread out and shifted to be more proton-rich
(see Figure \ref{fig:ye-histo}). The lowest original value of $Y_\mathrm{e}\approx 0.45$ 
is increased to $Y_\mathrm{e}\approx 0.48$. The higher values of $Y_\mathrm{e}$ are also 
correlated with higher values of entropy (see Figure \ref{fig:ye-s-plane}). 
Some of the higher-entropy tracers with the original values of $Y_\mathrm{e}\approx0.5$ 
now achieve the highest corrected values of $Y_\mathrm{e}\approx 0.58$.
Based on 2D simulations of similar progenitors, \citet{Wanajo.Mueller.ea:2018} 
also found the above correlation between $Y_\mathrm{e}$ and the asymptotic entropy of 
the inner ejecta. Our distribution of the corrected $Y_\mathrm{e}$ is also in qualitative agreement
with their distribution for the $11\,M_\odot$ model, which indicates that our
corrections of neutrino emission are reasonable. With or without these corrections,
most of the inner ejecta still have $Y_\mathrm{e}\approx 0.5$. This material is initially
at large radii and is then ejected by the shock with neither significant dissociation 
of its nuclei into free nucleons nor substantial exposure to neutrino irradiation. 
As a result, it essentially retains its initial $Y_\mathrm{e}$.

For tracers with peak temperatures below $6.5\,\mathrm{GK}$, NSE does not apply and the 
peak temperature is the key parameter for the nucleosynthesis.
Figure \ref{fig:tpeak_comparison} shows that the tracers in the 3D simulation 
get heated to higher temperatures 
compared to matter with the same initial radii in the 1D models. 
The tracer data also show a significant scatter, illustrating that 
no single 1D model could reproduce the range of conditions found in the 3D simulation.
As further illustration of this point, the 1D models do not include material that falls close to the PNS, 
reaches temperatures in excess of $10\,\mathrm{GK}$, and is still ejected eventually. 
For comparison with the 3D simulation, we also calculate the nucleosynthesis 
of the inner ejecta using the adopted 1D model. In addition to the above difference
in peak temperatures, matter ejected from near the PNS in the 3D simulation has more 
extreme values of $Y_\mathrm{e}$ than that in our adopted 1D model. Both these differences are 
among the key factors differentiating the nucleosynthesis results for the two types of 
models.

\subsection{Results for the Inner Ejecta}

\begin{figure*}
    \centering
    \includegraphics[width=\linewidth]{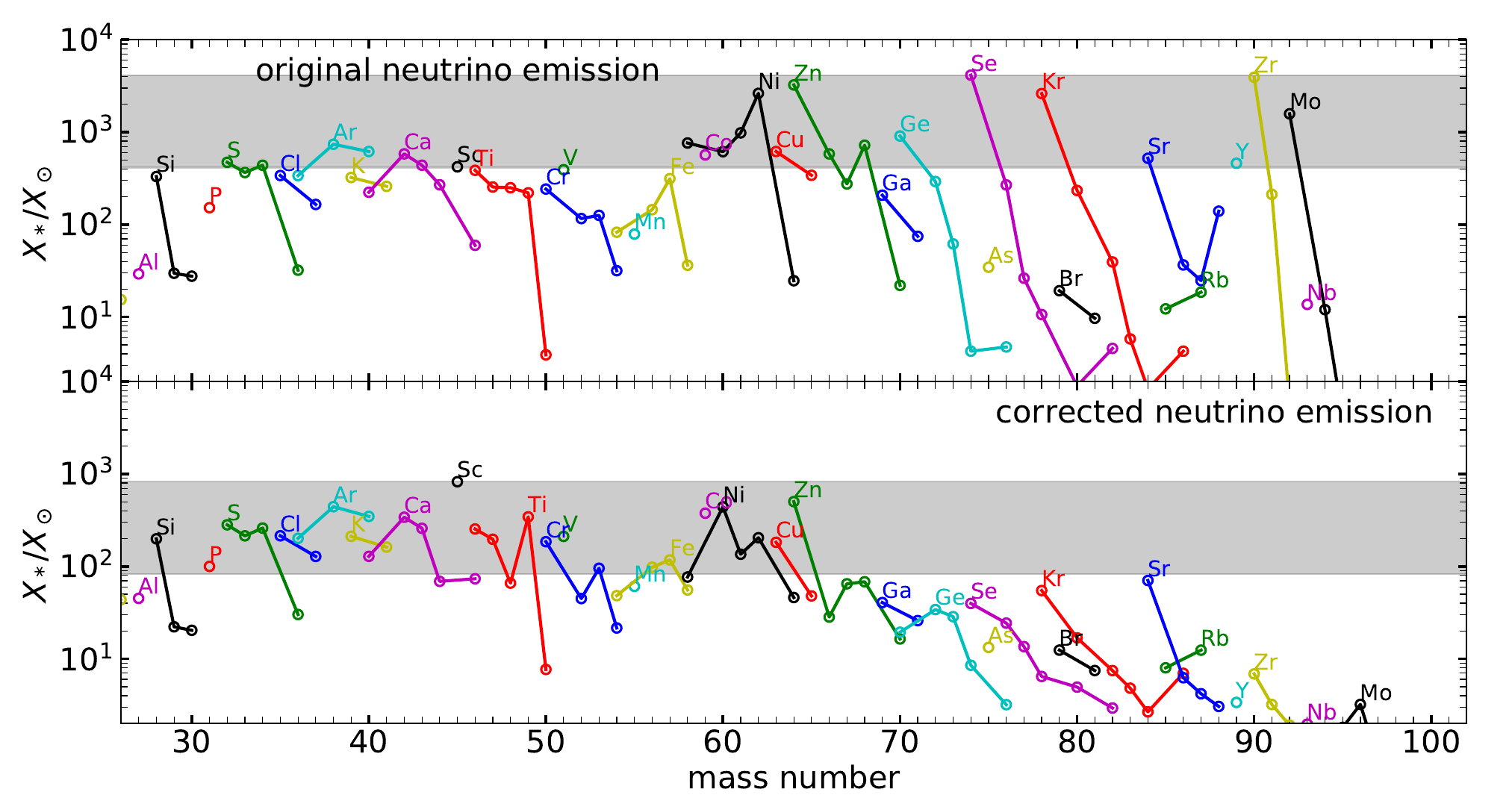}
    \caption{Production factors for stable isotopes made in the inner ejecta with the original 
    (top panel) and corrected (bottom panel) neutrino emission properties. The mass fraction $X_*$
    is relative to the total mass of $0.1\,M_\odot$ for the inner ejecta.
    The shaded bands include isotopes with production factors above 0.1 times the highest value.
    Isotopes of the same element are connected by line segments and colored for better visibility.  
    \label{fig:prod_factors_compare_lin} }
\end{figure*}

Figure \ref{fig:prod_factors_compare_lin} shows the final production factors $X_*/X_\odot$
for the stable isotopes after all decay is completed, where $X_*$ is the mass fraction 
of a stable isotope produced relative to the total mass ($M_{\rm in}=0.1\,M_\odot$) of the 
inner ejecta and $X_\odot$ is the corresponding solar mass 
fraction given by \citet{Asplund.Grevesse.ea:2009}. The shaded bands include isotopes
with production factors above 0.1 times the highest value. For these isotopes, $\geq 10\%$ 
of their solar inventory could be contributed by the CCSNe of concern. It is instructive
to compare the results calculated with the original (top panel) and corrected (bottom panel)
neutrino emission properties. The more neutron-rich conditions in the former case
(see Figure \ref{fig:ye-s-plane}) lead to the high production factors for the p-nuclei 
$^{74}$Se, $^{78}$Kr, and $^{92}$Mo, which is in agreement with the parametric studies of 
\citet{Hoffman.Woosley.ea:1996} and the calculations of \cite{Wanajo.Mueller.ea:2018} based on 
2D simulations for similar conditions. Here these isotopes are produced in
quasi-statistical equilibrium clusters for $Y_\mathrm{e}\approx 0.485$ and low values of entropy 
($\sim 10$--$20\,k_B$/baryon). The neutron-rich conditions are also responsible for the
high production factors for $^{62}$Ni, $^{64}$Zn, and $^{90}$Zr 
(see also \citealt{Hoffman.Woosley.ea:1996}).\footnote{The isotope $^{62}$Ni can be directly 
produced from NSE for $Y_\mathrm{e}\approx0.45$, which is close to its charge-to-mass ratio $Z/A=0.452$.
Most of its final yield, however, is produced as $^{62}$Zn with a half-life of $\approx 9\,\mathrm{h}$. 
With a higher value of $Z/A=0.484$, $^{62}$Zn is produced in a larger fraction of the material
undergoing complete Si burning and achieving NSE. The above two modes of $^{62}$Ni production were also
noted by \citet{Hoffman.Woosley.ea:1996}.}
As discussed above, however, we consider that the results calculated 
with the corrected neutrino emission properties are more realistic.
In this case, more tracers 
freeze out from NSE with $Y_\mathrm{e} \geq 0.5$ and higher values of entropy. Consequently, there is
no significant production for any of the p-nuclei mentioned above. Instead, the largest 
production factors are obtained for $^{45}$Sc, $^{60}$Ni, and $^{64}$Zn. In addition,
mostly because there are more $^4$He nuclei left at the end, the maximum production factor is 
smaller than that in the case with the original neutrino emission properties.
The yields from the tracers with the corrected neutrino emission properties are summarized in
Table~\ref{tab:yields_corrected}, which only includes stable isotopes up to mass number $A=100$ and
with yields exceeding $10^{-10}\,M_\odot$. The complete yields, including the mass fractions, are provided as supplemental material
in machine-readable format. 

 \begin{table*}[t]
 \caption{Yields\tablenotemark{a} of stable isotopes in $M_\odot$ from the tracers with the corrected neutrino emission properties
  \label{tab:yields_corrected}}
 \centering
 \begin{tabular}{ll|ll|ll|ll|ll} 
 \hline \hline
$^{1}$H & $1.55 \times 10^{-3} $ & $^{29}$Si & $1.01 \times 10^{-4} $ & $^{47}$Ti & $7.28 \times 10^{-6} $ & $^{65}$Cu & $1.39 \times 10^{-6} $ & $^{81}$Br & $6.05 \times 10^{-9} $ \\ 
$^{4}$He & $9.47 \times 10^{-3} $ & $^{30}$Si & $6.38 \times 10^{-5} $ & $^{48}$Ti & $2.47 \times 10^{-5} $ & $^{64}$Zn & $8.70 \times 10^{-5} $ & $^{78}$Kr & $3.59 \times 10^{-9} $ \\ 
$^{7}$Li & $5.68 \times 10^{-10} $ & $^{31}$P & $9.97 \times 10^{-5} $ & $^{49}$Ti & $9.62 \times 10^{-6} $ & $^{66}$Zn & $2.26 \times 10^{-6} $ & $^{80}$Kr & $7.16 \times 10^{-9} $ \\ 
$^{11}$B & $8.88 \times 10^{-10} $ & $^{32}$S & $1.57 \times 10^{-2} $ & $^{50}$Ti & $5.69 \times 10^{-8} $ & $^{67}$Zn & $7.94 \times 10^{-7} $ & $^{82}$Kr & $3.19 \times 10^{-9} $ \\ 
$^{12}$C & $8.12 \times 10^{-5} $ & $^{33}$S & $9.84 \times 10^{-5} $ & $^{51}$V & $1.28 \times 10^{-5} $ & $^{68}$Zn & $4.25 \times 10^{-6} $ & $^{83}$Kr & $3.04 \times 10^{-9} $ \\ 
$^{13}$C & $2.28 \times 10^{-8} $ & $^{34}$S & $6.93 \times 10^{-4} $ & $^{50}$Cr & $2.29 \times 10^{-5} $ & $^{70}$Zn & $3.03 \times 10^{-8} $ & $^{84}$Kr & $5.23 \times 10^{-9} $ \\ 
$^{14}$N & $3.62 \times 10^{-7} $ & $^{36}$S & $2.09 \times 10^{-7} $ & $^{52}$Cr & $1.11 \times 10^{-4} $ & $^{69}$Ga & $1.68 \times 10^{-7} $ & $^{86}$Kr & $7.74 \times 10^{-9} $ \\ 
$^{15}$N & $1.16 \times 10^{-6} $ & $^{35}$Cl & $1.25 \times 10^{-4} $ & $^{53}$Cr & $2.73 \times 10^{-5} $ & $^{71}$Ga & $6.02 \times 10^{-8} $ & $^{85}$Rb & $4.42 \times 10^{-9} $ \\ 
$^{16}$O & $1.87 \times 10^{-2} $ & $^{37}$Cl & $2.36 \times 10^{-5} $ & $^{54}$Cr & $1.05 \times 10^{-6} $ & $^{70}$Ge & $5.10 \times 10^{-8} $ & $^{87}$Rb & $2.88 \times 10^{-9} $ \\ 
$^{17}$O & $2.73 \times 10^{-7} $ & $^{36}$Ar & $2.00 \times 10^{-3} $ & $^{55}$Mn & $1.31 \times 10^{-4} $ & $^{72}$Ge & $2.54 \times 10^{-7} $ & $^{84}$Sr & $3.22 \times 10^{-9} $ \\ 
$^{18}$O & $2.36 \times 10^{-8} $ & $^{38}$Ar & $8.41 \times 10^{-4} $ & $^{54}$Fe & $5.60 \times 10^{-4} $ & $^{73}$Ge & $4.11 \times 10^{-8} $ & $^{86}$Sr & $2.46 \times 10^{-9} $ \\ 
$^{19}$F & $1.04 \times 10^{-7} $ & $^{40}$Ar & $9.63 \times 10^{-7} $ & $^{56}$Fe & $1.85 \times 10^{-2} $ & $^{74}$Ge & $3.19 \times 10^{-8} $ & $^{87}$Sr & $7.39 \times 10^{-10} $ \\ 
$^{20}$Ne & $3.64 \times 10^{-3} $ & $^{39}$K & $1.20 \times 10^{-4} $ & $^{57}$Fe & $5.14 \times 10^{-4} $ & $^{76}$Ge & $8.03 \times 10^{-9} $ & $^{88}$Sr & $5.43 \times 10^{-9} $ \\ 
$^{21}$Ne & $3.80 \times 10^{-6} $ & $^{41}$K & $6.88 \times 10^{-6} $ & $^{58}$Fe & $7.17 \times 10^{-6} $ & $^{75}$As & $1.44 \times 10^{-8} $ & $^{89}$Y & $1.70 \times 10^{-9} $ \\ 
$^{22}$Ne & $1.30 \times 10^{-6} $ & $^{40}$Ca & $1.26 \times 10^{-3} $ & $^{59}$Co & $2.04 \times 10^{-4} $ & $^{74}$Se & $7.42 \times 10^{-9} $ & $^{90}$Zr & $1.20 \times 10^{-8} $ \\ 
$^{23}$Na & $4.70 \times 10^{-5} $ & $^{42}$Ca & $2.33 \times 10^{-5} $ & $^{58}$Ni & $6.27 \times 10^{-4} $ & $^{76}$Se & $2.93 \times 10^{-8} $ & $^{91}$Zr & $6.23 \times 10^{-10} $ \\ 
$^{24}$Mg & $8.90 \times 10^{-4} $ & $^{43}$Ca & $3.75 \times 10^{-6} $ & $^{60}$Ni & $1.43 \times 10^{-3} $ & $^{77}$Se & $1.01 \times 10^{-8} $ & $^{92}$Zr & $3.01 \times 10^{-10} $ \\ 
$^{25}$Mg & $1.18 \times 10^{-4} $ & $^{44}$Ca & $1.59 \times 10^{-5} $ & $^{61}$Ni & $1.48 \times 10^{-5} $ & $^{78}$Se & $5.59 \times 10^{-9} $ & $^{94}$Zr & $3.10 \times 10^{-10} $ \\ 
$^{26}$Mg & $1.16 \times 10^{-4} $ & $^{46}$Ca & $1.50 \times 10^{-8} $ & $^{62}$Ni & $9.02 \times 10^{-5} $ & $^{80}$Se & $1.17 \times 10^{-8} $ & $^{93}$Nb & $1.87 \times 10^{-10} $ \\ 
$^{27}$Al & $1.66 \times 10^{-4} $ & $^{45}$Sc & $5.39 \times 10^{-6} $ & $^{64}$Ni & $1.48 \times 10^{-6} $ & $^{82}$Se & $5.52 \times 10^{-9} $ & $^{92}$Mo & $1.55 \times 10^{-10} $ \\ 
$^{28}$Si & $2.14 \times 10^{-2} $ & $^{46}$Ti & $1.02 \times 10^{-5} $ & $^{63}$Cu & $1.33 \times 10^{-5} $ & $^{79}$Br & $6.79 \times 10^{-9} $ & $^{96}$Mo & $1.34 \times 10^{-10} $ \\ 
\hline
\end{tabular} 
\tablenotetext{a}{Yields less than $10^{-10}\,M_\odot$ and those for $A>100$ are not shown. A complete table is available as supplemental material.}
 \end{table*}
 
The high production factor for $^{45}$Sc in proton-rich ejecta resembles
the previous results of, e.g., \citet{Pruet.Woosley.ea:2005} and \citet{Frohlich.Hauser.ea:2006}.  
Starting mostly from $^{40}$Ca, which is abundant at the freeze-out from NSE, 
a sequence of proton captures produces highly neutron-deficient nuclei 
up to $^{45}$Cr. Due to the very low proton separation energy of $^{46}$Mn, 
the reaction flow needs to wait for the $\beta$-decay of $^{45}$Cr
with a half-life of $64\,\mathrm{ms}$. Because matter cools down considerably on 
this timescale, charged-particle reactions quickly freeze out and the decay chain of
$^{45}$Cr eventually produces $^{45}$Sc. In addition, sufficiently fast expansion
is required to prevent proton captures on $^{45}$V along the decay chain so that
a large production factor of $^{45}$Sc can be obtained. 
Such expansion is found for the higher-entropy ejecta in 
multidimensional simulations. Unfortunately, the longest half-life along the decay chain to 
$^{45}$Sc is only $\,\approx 3\,\mathrm{h}$ for $^{45}$Ti, which is too short to allow 
late-time observations. 

The production of $^{64}$Zn depends strongly on $Y_\mathrm{e}$. 
For tracers freezing out from NSE with $Y_\mathrm{e}\approx 0.47$--0.48,
$^{64}$Zn is produced almost directly from NSE.
For proton-rich conditions, $^{64}$Zn is made as $^{64}$Ga and $^{64}$Ge 
by a sequence of proton capture reactions following the particle-rich freeze-out.
Relatively large production factors for $^{64}$Zn were also found by
\citet{Wanajo.Mueller.ea:2018}. 
Because the production of $^{64}$Zn is dominated by tracers achieving NSE,
it is insensitive to the initial composition of the progenitor.
Consequently, the large production factor of $^{64}$Zn for our progenitor model 
with an initial solar metallicity may also help explain the observations of 
[Zn/Fe]~$=\log({\rm Zn/Fe})-\log({\rm Zn/Fe})_\odot$ in metal-poor stars.
\citet{Ezzeddine.Frebel.ea:2019} recently measured [Zn/Fe]~$=0.8\pm 0.25$ 
for an ultra-metal-poor star and had difficulty reconciling this result with 
the predictions of 1D CCSN models. Based on the 3D simulation, we find
that our model can provide [Zn/Fe]~$=1.11$ and $0.43$ with the original 
and corrected neutrino emission properties, respectively. In contrast,
our adopted 1D model gives a much lower value of [Zn/Fe]~$=0.10$.
Our results indicate that multidimensional dynamics and accurate neutrino emission
properties are crucial to understanding the particular measurement of 
\citet{Ezzeddine.Frebel.ea:2019} and the general observations of [Zn/Fe] 
in metal-poor stars. The production of Zn and Fe by metal-free and metal-poor
CCSNe certainly merits detailed investigation with 3D simulations and
accurate neutrino transport.

The inner ejecta also produce $^{56}$Ni and $^{44}$Ti,
which are important due to the observational signatures from their radioactivity.
Our tracer calculations give $0.016\, M_\odot$ and $0.018\, M_\odot$ of $^{56}$Ni
with the original and corrected neutrino emission properties, respectively.
By comparison, our adopted 1D model gives $\approx 0.023\,M_\odot$, which is
insensitive to the neutrino emission properties. All of the above values\footnote{
The mass resolution is typically $4\times 10^{-3}\,M_\odot$ for the 1D model, :sp :::dsdfeeeef
which is much coarser than that of the tracers. This difference results in 
different sampling of nucleosynthesis conditions, which may also slightly 
affect the yields.}
are in approximate agreement with what is expected from similar low-mass CCSNe 
\citep{Mueller.ea:2016,Sukhbold.Ertl.ea:2016,Ebinger.Curtis.ea:2019}
and with the observed lower $^{56}$Ni production by less energetic events
\citep{Hamuy:2003,Pejcha.Prieto:2015,Mueller.Pejcha.ea:2017}.
The larger $^{56}$Ni yield in our adopted 1D model can be understood as follows.
Although the amount of matter reaching temperatures above $6.5\,\mathrm{GK}$
is $0.028\,M_\odot$ for the 3D simulation and only $0.014\,M_\odot$ for the 1D model,
the amount of $^{56}$Ni produced from this NSE component undergoing
complete Si burning is almost the same for both models. This result can be
traced to the more extreme values of $Y_\mathrm{e}$ and the much larger mass fraction 
of $\alpha$ particles at the freeze-out of the 3D tracers, both of which
reduce the production of $^{56}$Ni. In addition, the amount of matter
reaching temperatures of $\sim 3$--$6.5\,\mathrm{GK}$ in the 3D simulation is smaller
than that in the 1D model. The extra production of $^{56}$Ni for the latter
mostly originates in this component undergoing incomplete Si burning.
Note that the $^{56}$Ni yields from 1D calculations are
particularly sensitive to the prescribed explosion 
and to the choice of parameters \citep{Young.Fryer:2007}.
For example, another 1D model with $M_\mathrm{cut}=1.38 M_\odot$ gives 
a $^{56}$Ni yield of $0.015 M_\odot$. In principle, a suitable average of 1D models 
may reproduce the $^{56}$Ni yield from the 3D simulation. The same reproduction, however,
cannot be achieved for those isotopes that are sensitive to the value of $Y_\mathrm{e}$ and hence
exhibit the most significant differences between 1D and 3D models (see
\S\ref{sec:total_yields}).

The production of $^{44}$Ti is generally sensitive to a combination of factors.
The tracer calculations with the original neutrino emission properties give 
$3.73 \times 10^{-5}\,M_\odot$ of $^{44}$Ti,
to be compared with $1.13 \times 10^{-5}\,M_\odot$ for our adopted 1D model
(other 1D models with $M_\mathrm{cut}=1.31$ and $1.38\,M_\odot$ give
$0.98 \times 10^{-5}$ and $1.2 \times 10^{-5}\,M_\odot$ of $^{44}$Ti, respectively).
This result can be explained
by the higher values of entropy and the more particle-rich freeze-out for the
tracers, both of which help the $^{44}$Ti production. With the corrected neutrino 
emission properties, however, the higher-entropy tracers are more proton-rich, 
which increases the production of $^{45}$Sc and $^{64}$Zn at the expense of $^{44}$Ti.
As a result, the $^{44}$Ti yield of $1.58\times 10^{-5}\,M_\odot$ is only slightly 
above that for the 1D model, which is again insensitive to the neutrino emission
properties. Therefore, while the higher values of entropy found in 3D simulations 
generally tend to increase the $^{44}$Ti yield, the more extreme values of $Y_\mathrm{e}$ 
found in such models can also have a large impact. The resulting uncertainty in
the $^{44}$Ti yield can be removed only by 3D simulations with accurate neutrino 
transport. Such simulations are also required to address whether the 3D explosion
of an appropriate progenitor can account for the rather high $^{44}$Ti yield of 
$\sim (0.3$--$3.9)\times 10^{-4}\,M_\odot$ inferred for SN~1987A 
\citep{Grebenev.Lutovinov.ea:2012,Seitenzahl.Timmes.ea:2014,Boggs.Harrison.ea:2015}.

\section{Nucleosynthesis of the Outer Ejecta and Total CCSN Yields}
\label{sec:nucleo_outer}

As discussed in \S\ref{sec:nucleo_general}, we use our adopted 1D model with 
$M_{\mathrm{cut}}=1.35 M_\odot$ to
calculate the nucleosynthesis of the outer ejecta, which are separated from the
inner ejecta at an enclosed mass of $M_{\rm tr}=1.45\,M_\odot$.
The separation between the tracer particles and the outer ejecta is located in 
the O/Ne shell and has a peak temperature of 
$T_{\mathrm{peak}}=2.2$ GK in the 1D model, which fits well with our lower limit 
of $T_{\mathrm{peak}}=2$ GK for the selected tracers representing the inner 
ejecta.\footnote{A small amount of material with $T_{\mathrm{peak}}=2.0$--2.2 GK
is represented by both tracers and the 1D model. This overlap is consistent with the 
scatter of tracers at $T_{\mathrm{peak}}\sim 2$ GK in Figure \ref{fig:tpeak_comparison}.}
We also have done calculations using other 1D models with $M_{\mathrm{cut}}=1.31$ 
and $1.38 M_\odot$ and found no significant difference in the yields because the conditions 
in the outer ejecta are almost identical for $M_{\mathrm{cut}}=1.31$--$1.38 M_\odot$
(see Figure \ref{fig:tpeak_comparison}). The results from our adopted 1D model are
presented below.

The $\nu$ process, i.e., the production of isotopes in the outer ejecta due to 
neutrino-nucleus reactions \citep{Woosley.Hartmann.ea:1990}, leads to important nucleosynthetic signatures, including 
the production of the SLR $^{10}$Be. This process depends on both the neutrino emission 
from the PNS and the shock propagation through the outer ejecta, thereby providing
a direct link between the inner evolution described by the 3D simulation and
the outer evolution described by our adopted 1D model. We use the corrected neutrino
emission properties to calculate the rates of neutrino-nucleus reactions for 
the $\nu$ process. \citet{Sieverding.Langanke.ea:2019} found that the early phase 
of neutrino emission is important for this process. This phase is automatically
included in our calculations.

\subsection{Total CCSN Yields}
\label{sec:total_yields}
Combining the results for the inner and outer ejecta with the corrected neutrino
emission properties, we give the total yields of 
the CCSN in Table~\ref{tab:yields_combined_corrected}, which only includes
stable isotopes up to $A=100$ and with yields exceeding $10^{-10}\,M_\odot$.
The complete table is provided as supplemental material in machine-readable format. 
The contributions from the outer ejecta can be obtained from comparing
Tables~\ref{tab:yields_corrected} and \ref{tab:yields_combined_corrected}.

 \begin{table*}[t]
  \caption{Total CCSN yields\tablenotemark{a} of stable isotopes in $M_\odot$ with the corrected neutrino emission properties
   \label{tab:yields_combined_corrected}}
 \centering
\begin{tabular}{ll|ll|ll|ll|ll} 
\hline \hline
$^{1}$H & $5.31 $ & $^{28}$Si & $2.75 \times 10^{-2} $ & $^{49}$Ti & $1.13 \times 10^{-5} $ & $^{70}$Zn & $1.93 \times 10^{-7} $ & $^{87}$Rb & $5.90 \times 10^{-8} $ \\ 
$^{2}$H & $1.23 \times 10^{-10} $ & $^{29}$Si & $4.78 \times 10^{-4} $ & $^{50}$Ti & $1.91 \times 10^{-6} $ & $^{69}$Ga & $7.89 \times 10^{-7} $ & $^{84}$Sr & $5.60 \times 10^{-9} $ \\ 
$^{3}$He & $3.22 \times 10^{-4} $ & $^{30}$Si & $3.13 \times 10^{-4} $ & $^{51}$V & $1.60 \times 10^{-5} $ & $^{71}$Ga & $4.82 \times 10^{-7} $ & $^{86}$Sr & $5.71 \times 10^{-8} $ \\ 
$^{4}$He & $3.33 $ & $^{31}$P & $1.79 \times 10^{-4} $ & $^{50}$Cr & $2.93 \times 10^{-5} $ & $^{70}$Ge & $7.14 \times 10^{-7} $ & $^{87}$Sr & $3.71 \times 10^{-8} $ \\ 
$^{7}$Li & $1.31 \times 10^{-7} $ & $^{32}$S & $1.86 \times 10^{-2} $ & $^{52}$Cr & $2.41 \times 10^{-4} $ & $^{72}$Ge & $1.03 \times 10^{-6} $ & $^{88}$Sr & $4.25 \times 10^{-7} $ \\ 
$^{9}$Be & $1.42 \times 10^{-10} $ & $^{33}$S & $1.24 \times 10^{-4} $ & $^{53}$Cr & $4.25 \times 10^{-5} $ & $^{73}$Ge & $2.95 \times 10^{-7} $ & $^{89}$Y & $1.02 \times 10^{-7} $ \\ 
$^{10}$B & $1.56 \times 10^{-9} $ & $^{34}$S & $8.38 \times 10^{-4} $ & $^{54}$Cr & $6.29 \times 10^{-6} $ & $^{74}$Ge & $9.94 \times 10^{-7} $ & $^{90}$Zr & $1.24 \times 10^{-7} $ \\ 
$^{11}$B & $3.23 \times 10^{-7} $ & $^{36}$S & $1.43 \times 10^{-6} $ & $^{55}$Mn & $2.47 \times 10^{-4} $ & $^{76}$Ge & $1.63 \times 10^{-7} $ & $^{91}$Zr & $2.65 \times 10^{-8} $ \\ 
$^{12}$C & $5.77 \times 10^{-2} $ & $^{35}$Cl & $1.56 \times 10^{-4} $ & $^{54}$Fe & $1.16 \times 10^{-3} $ & $^{75}$As & $1.51 \times 10^{-7} $ & $^{92}$Zr & $3.96 \times 10^{-8} $ \\ 
$^{13}$C & $6.98 \times 10^{-4} $ & $^{37}$Cl & $3.95 \times 10^{-5} $ & $^{56}$Fe & $2.84 \times 10^{-2} $ & $^{74}$Se & $1.70 \times 10^{-8} $ & $^{94}$Zr & $4.02 \times 10^{-8} $ \\ 
$^{14}$N & $2.68 \times 10^{-2} $ & $^{36}$Ar & $2.52 \times 10^{-3} $ & $^{57}$Fe & $7.78 \times 10^{-4} $ & $^{76}$Se & $2.03 \times 10^{-7} $ & $^{93}$Nb & $1.74 \times 10^{-8} $ \\ 
$^{15}$N & $2.03 \times 10^{-5} $ & $^{38}$Ar & $9.47 \times 10^{-4} $ & $^{58}$Fe & $1.06 \times 10^{-4} $ & $^{77}$Se & $1.32 \times 10^{-7} $ & $^{92}$Mo & $8.02 \times 10^{-9} $ \\ 
$^{16}$O & $1.66 \times 10^{-1} $ & $^{40}$Ar & $1.71 \times 10^{-6} $ & $^{59}$Co & $2.61 \times 10^{-4} $ & $^{78}$Se & $3.73 \times 10^{-7} $ & $^{94}$Mo & $5.41 \times 10^{-9} $ \\ 
$^{17}$O & $5.34 \times 10^{-5} $ & $^{39}$K & $1.51 \times 10^{-4} $ & $^{58}$Ni & $1.05 \times 10^{-3} $ & $^{80}$Se & $7.46 \times 10^{-7} $ & $^{95}$Mo & $9.47 \times 10^{-9} $ \\ 
$^{18}$O & $7.44 \times 10^{-4} $ & $^{41}$K & $9.81 \times 10^{-6} $ & $^{60}$Ni & $1.62 \times 10^{-3} $ & $^{82}$Se & $1.16 \times 10^{-7} $ & $^{96}$Mo & $1.71 \times 10^{-8} $ \\ 
$^{19}$F & $7.21 \times 10^{-6} $ & $^{40}$Ca & $1.78 \times 10^{-3} $ & $^{61}$Ni & $3.35 \times 10^{-5} $ & $^{79}$Br & $1.43 \times 10^{-7} $ & $^{97}$Mo & $5.68 \times 10^{-9} $ \\ 
$^{20}$Ne & $4.54 \times 10^{-2} $ & $^{42}$Ca & $2.73 \times 10^{-5} $ & $^{62}$Ni & $1.29 \times 10^{-4} $ & $^{81}$Br & $1.25 \times 10^{-7} $ & $^{98}$Mo & $1.47 \times 10^{-8} $ \\ 
$^{21}$Ne & $1.28 \times 10^{-4} $ & $^{43}$Ca & $4.67 \times 10^{-6} $ & $^{64}$Ni & $2.22 \times 10^{-5} $ & $^{78}$Kr & $6.93 \times 10^{-9} $ & $^{100}$Mo & $5.81 \times 10^{-9} $ \\ 
$^{22}$Ne & $3.16 \times 10^{-3} $ & $^{44}$Ca & $2.86 \times 10^{-5} $ & $^{63}$Cu & $3.21 \times 10^{-5} $ & $^{80}$Kr & $3.16 \times 10^{-8} $ & $^{96}$Ru & $2.08 \times 10^{-9} $ \\ 
$^{23}$Na & $1.59 \times 10^{-3} $ & $^{46}$Ca & $8.20 \times 10^{-8} $ & $^{65}$Cu & $6.51 \times 10^{-6} $ & $^{82}$Kr & $1.67 \times 10^{-7} $ & $^{98}$Ru & $7.23 \times 10^{-10} $ \\ 
$^{24}$Mg & $9.54 \times 10^{-3} $ & $^{45}$Sc & $5.86 \times 10^{-6} $ & $^{64}$Zn & $9.63 \times 10^{-5} $ & $^{83}$Kr & $1.39 \times 10^{-7} $ & $^{99}$Ru & $4.97 \times 10^{-9} $ \\ 
$^{25}$Mg & $1.81 \times 10^{-3} $ & $^{46}$Ti & $1.24 \times 10^{-5} $ & $^{66}$Zn & $9.69 \times 10^{-6} $ & $^{84}$Kr & $6.63 \times 10^{-7} $ & $^{100}$Ru & $5.25 \times 10^{-9} $ \\ 
$^{26}$Mg & $1.89 \times 10^{-3} $ & $^{47}$Ti & $9.26 \times 10^{-6} $ & $^{67}$Zn & $2.19 \times 10^{-6} $ & $^{86}$Kr & $2.47 \times 10^{-7} $ &  \\ 
$^{27}$Al & $1.15 \times 10^{-3} $ & $^{48}$Ti & $4.56 \times 10^{-5} $ & $^{68}$Zn & $9.93 \times 10^{-6} $ & $^{85}$Rb & $1.33 \times 10^{-7} $ &  \\ 
\hline
\end{tabular} 
\tablenotetext{a}{Yields less than $10^{-10}\,M_\odot$ or for $A>100$ are not shown. A complete table is available as supplemental material.}
 \end{table*}
 
\begin{figure*}
    \centering
    \includegraphics[width=\textwidth]{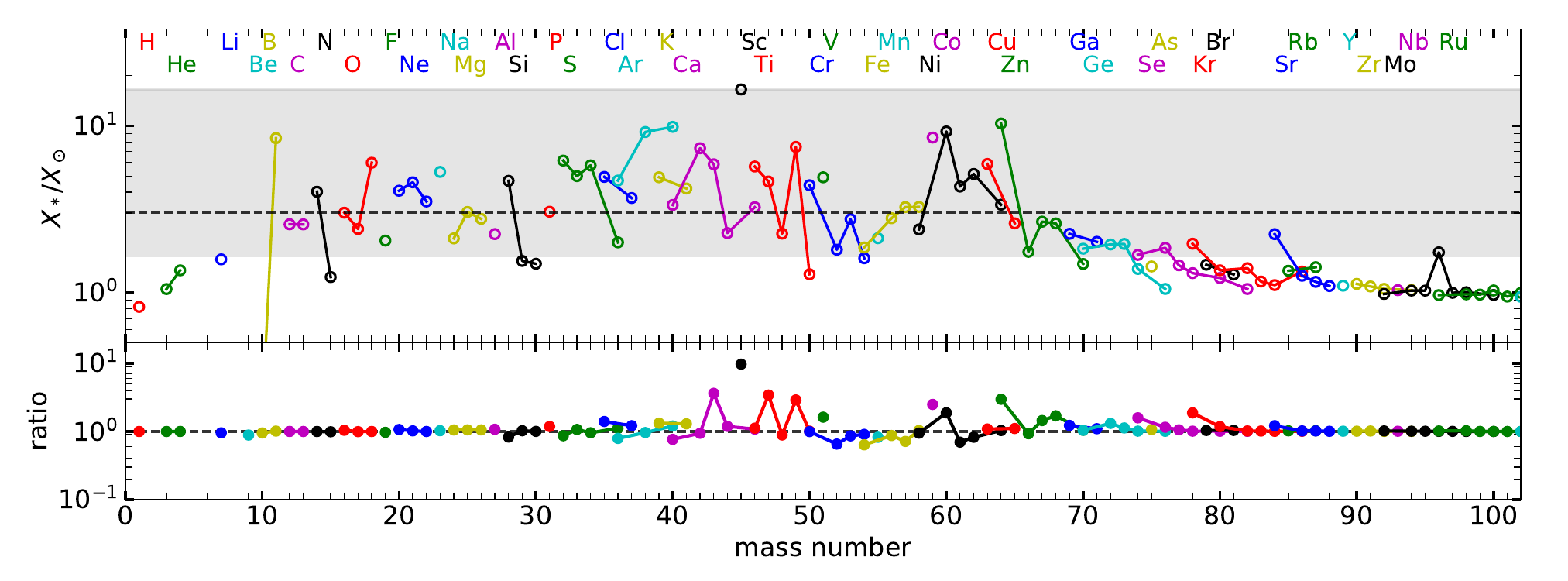}
    \caption{Production factors for stable isotopes in the entire CCSN ejecta with the 
    corrected neutrino emission properties (top panel) and the ratios of yields from
    our combined model to those from the adopted 1D model (bottom panel). The mass 
    fraction $X_*$ is relative to the total ejecta mass of $9.04\,M_\odot$. 
    The shaded band includes isotopes with production factors above 0.1 times 
    the highest value. The dashed line indicates the production factor of $^{16}$O.
    Isotopes of the same element are connected by line segments and colored for better 
    visibility.}
    \label{fig:yield_comparison}
\end{figure*}

The top panel of Figure~\ref{fig:yield_comparison} shows the production factors of
the stable isotopes in the entire CCSN ejecta, where the mass fraction of each isotope 
is relative to the total ejecta mass of $9.04\,M_\odot$.\footnote{Note that a total of
$1.41\,M_\odot$ of the H envelope are lost through winds during the pre-CCSN evolution.}
The largest production factor 
of 16.5 corresponds to $^{45}$Sc. The isotopes with production factors exceeding half of 
this value are $^{11}$B, $^{38,40}$Ar, $^{59}$Co, $^{60}$Ni, and $^{64}$Zn.
The shaded band indicates production factors above 0.1 times the highest value and
includes the majority of the stable isotopes between B and Ge. These results suggest
that had $11.8\,M_\odot$ CCSNe made all of the $^{45}$Sc in the solar inventory, they
would have also contributed $\gtrsim 10\%$ of the majority of other stable isotopes 
between B and Ge. In particular, with a production factor of 3.01 for $^{16}$O, they 
would have contributed $\approx 18\%$ of this important isotope. The top panel of
Figure~\ref{fig:yield_comparison} also shows that except for $^{96}$Mo, the stable 
isotopes with $A\sim 90$--100 have production factors close to unity. These isotopes 
simply reflect the initial composition of the ejecta. This result also extends to
nearly all of those stable isotopes with $A>100$ (not shown) because the corresponding 
initial composition is essentially unaffected by the pre-CCSN evolution of the progenitor
or the CCSN process. The small number of exceptions are the isotopes produced by the $\gamma$
process (e.g., $^{152}$Gd) and the $\nu$ process (e.g., $^{138}$La and $^{180}$Ta)
during the shock propagation.

The isotope with the largest production factor, $^{45}$Sc, is produced in the inner
ejecta. Other major isotopes produced by the inner ejecta can be identified by comparing
the shaded band in Figure~\ref{fig:yield_comparison} with that in the bottom panel of 
Figure~\ref{fig:prod_factors_compare_lin} (note that the mass fraction in the latter
figure is relative to the total inner ejecta mass of $0.1\,M_\odot$, but this 
normalization does not affect the placement of the shaded band). To see the effects
of the 3D explosion, we show in the bottom panel of 
Figure~\ref{fig:yield_comparison} the ratios of yields from our combined model to 
those from the adopted 1D model. 
Isotopes that are mostly contributed by the outer ejecta 
are unaffected by the difference in the explosion. These isotopes include the elements 
up to Si. The largest change due to the 3D explosion is the increase of the $^{45}$Sc 
yield by an order of magnitude. The yields of $^{43}$Ca and $^{47,49}$Ti increase 
by factors of 3--4. With a large amount of matter freezing out from the proton-rich NSE 
in the 3D simulation, the yield pattern of the Ni isotopes changes and the $^{60}$Ni 
yield increases by more than a factor of $3$. The yield of $^{59}$Co increases by a factor 
of 2.3 under the same conditions. In addition, the 3D proton-rich component of the inner ejecta 
produces $\sim 2$--3 times more of $^{64}$Zn and the light $p$-isotopes $^{74}$Se and $^{78}$Kr.
The above results can be largely traced to the more proton-rich conditions of the inner ejecta 
due to more neutrino processing in 3D. 
Other 1D models with $M_{\rm{cut}}=1.31$ and $1.38\,M_\odot$ show
similar differences from the 3D model. In general, because the complex history of neutrino 
heating of the tracers is hard to replicate with 1D models, it is unlikely that 
the overall abundance pattern calculated using the tracers can be reproduced by
a superposition of 1D models.

\subsection{Uncertainties of $^{10}$Be Production}
\label{sec:be10}
In addition to the stable isotopes, our combined model gives the yields of
SLRs, which are used in \S\ref{sec:SLRs} to explore the $11.8\,M_\odot$ 
CCSN as a candidate trigger for the formation of the solar system
\citep{Banerjee.Qian.ea:2016}.
In this subsection, we discuss the production of the SLR $^{10}$Be and the
associated uncertainties. As suggested by \citet{Banerjee.Qian.ea:2016}, CCSNe 
might be a significant source of $^{10}$Be in the interstellar medium (ISM),
which contradicts the common assumption that this SLR can only be produced 
by high-energy collisions between nuclei. Therefore, our discussion has other 
implications even if the $11.8\,M_\odot$ CCSN might not have been the trigger 
for solar system formation.

The CCSN production of $^{10}$Be proceeds predominantly through the neutral-current (NC) 
reaction $^{12}$C$(\nu,\nu ' \mathrm{pp})^{10}$Be and the charged-current (CC) reactions
$^{12}$C$(\bar{\nu}_\mathrm{e}, \mathrm{e}^+ \mathrm{np})^{10}$Be and $^{11}$C$(\bar{\nu}_\mathrm{e}, \mathrm{e}^+ \mathrm{p})^{10}$Be.
The role of $^{12}$C$(\bar{\nu}_\mathrm{e}, \mathrm{e}^+ \mathrm{np})^{10}$Be was noted by \citet{Takigawa.Miki.ea:2008}.
All of these reactions occur in the relatively thin C/O shell of the 
$11.8\,M_\odot$ CCSN, where the peak post-shock temperature is just below
$1\,\mathrm{GK}$. This material is part of the outer ejecta and is not affected
by the details of the explosion so long as an appropriate explosion energy is used.
We adopt the cross sections of \citet{Yoshida.Suzuki.ea:2008} based on
shell-model calculations for $^{12}$C$(\nu,\nu ' \mathrm{pp})^{10}$Be and 
$^{12}$C$(\bar{\nu}_\mathrm{e}, \mathrm{e}^+ \mathrm{np})^{10}$Be, and the cross section of 
\citet{Sieverding.Martinez.ea:2018} based on random-phase approximation
for $^{11}$C$(\bar{\nu}_\mathrm{e}, \mathrm{e}^+ \mathrm{p})^{10}$Be. The $\bar\nu_\mathrm{e}$-induced CC reactions
typically have larger cross sections than the NC reaction induced by all neutrino
species. As a result, the NC and CC channels make approximately the same contributions
to the production of $^{10}$Be. In contrast, the $\nu_\mathrm{e}$-induced CC reaction
$^{10}$Be$(\nu_\mathrm{e},\mathrm{e}^-)^{10}$B destroys $^{10}$Be. The cross section of this reaction
is taken from \citet{Sieverding.Martinez.ea:2018}.
The destruction of $^{10}$Be by $^{10}$Be$(p,\alpha)^{7}$Li can also be important.
By default, we use the rate in the JINA REACLIB library \citep{Cyburt.Amthor.ea:2010} 
for this reaction, which is based on the estimates of \citet{Wagoner:1969} from the 
statistical model. 

The nominal yield of $^{10}$Be from the $11.8\,M_\odot$ CCSN is $6.08\times 10^{-11}\,M_\odot$
(see the first entry in Table~\ref{tab:be10_sensitivity}). There are, however, large 
uncertainties in this yield. One factor is the large uncertainty in the rate of
$^{10}$Be$(\mathrm{p},\alpha)^{7}$Li \citep{Iliadis.Longland.ea:2010}. For example, we have repeated 
the calculations leaving out this reaction but with no other changes and found a factor of 
$\approx 3$ increase in the $^{10}$Be yield (see Table~\ref{tab:be10_sensitivity}).
Further uncertainties in this yield come from the possibility of neutrino flavor 
oscillations that could significantly alter the effective neutrino energy spectra
for the production and destruction of $^{10}$Be. Whereas flavor oscillations do not
affect the NC production, the CC production (destruction) could be significantly
enhanced if $\bar{\nu}_x$ ($\nu_x$) with harder spectra (see Figure~\ref{fig:nu_signal})
were converted into $\bar{\nu}_\mathrm{e}$ ($\nu_\mathrm{e}$). Similar effects have been discussed previously for 
the $\nu$ process in general \citep[e.g.,][]{Yoshida.Kajino.ea:2006,Wu.Qian.ea:2015}. 
For illustration, we consider two extreme scenarios where either
$\bar{\nu}_\mathrm{e}\leftrightarrow\bar{\nu}_x$ or $\nu_\mathrm{e}\leftrightarrow\nu_x$ oscillations 
cause the corresponding neutrino species to exchange their spectra, thereby having 
the largest effect on the $^{10}$Be production or destruction. As shown in
Table~\ref{tab:be10_sensitivity}, when $^{10}$Be$(p,\alpha)^{7}$Li is (not)
included in the calculations, $\bar{\nu}_\mathrm{e}\leftrightarrow\bar{\nu}_x$ 
oscillations can increase the $^{10}$Be yield by a factor of $\approx 3.3$ (2.8), whereas
$\nu_\mathrm{e}\leftrightarrow\nu_x$ oscillations can decrease it by a factor of $\approx 1.7$ (1.2).

\begin{table}
 \caption{Uncertainties in the $^{10}$Be yield}
     \centering
     \begin{tabular}{ccc}\hline\hline
          $^{10}$Be$(p,\alpha)^7$Li
         & Flavor Oscillations & $^{10}$Be Yield ($10^{-10} M_\odot$)\\ \hline
           default   & no & 0.608 \\
           off    & no & 1.80 \\
           default  & $\bar{\nu}_\mathrm{e}\leftrightarrow\bar{\nu}_x$ & 1.99 \\
           off    & $\bar{\nu}_\mathrm{e}\leftrightarrow\bar{\nu}_x$ & 5.06 \\
           default & $\nu_\mathrm{e}\leftrightarrow\nu_x$ & 0.348 \\
            off    & $\nu_\mathrm{e}\leftrightarrow\nu_x$ & 1.48 \\
           \hline
     \end{tabular}
     \label{tab:be10_sensitivity}
\end{table}

Without neutrino flavor oscillations,
our nominal $^{10}$Be yield from the $11.8\,M_\odot$ CCSN is about one-fith 
the value of $3.26\times10^{-10}\,M_\odot$ obtained by \citet{Banerjee.Qian.ea:2016}.
Compared to our calculations,
that work used a lower explosion energy of $0.1\times 10^{50}$~erg and somewhat
different neutrino spectra. The main differences, however, are that
it ignored $^{10}$Be$(p,\alpha)^{7}$Li, $^{10}$Be$(\nu_\mathrm{e},e^-)^{10}$B, and
the $\bar\nu_\mathrm{e}$-induced CC production of $^{10}$Be. The destruction of $^{10}$Be by 
the first two reactions overwhelms the additional production by $\bar\nu_\mathrm{e}$, which
explains our much smaller nominal yield. 

\citet{Banerjee.Qian.ea:2016} also calculated $^{10}$Be yields for more massive CCSNe.
Those yields should be checked by detailed calculations including those reactions 
ignored by that work. The effects of neutrino spectra and flavor oscillations should 
be explored as well. Such efforts are required to quantify CCSNe as a source of 
$^{10}$Be in the ISM. 

\section{Overall Production of SLRs}
\label{sec:SLRs}
\begin{table*}[ht]
\caption{Yields of SLRs and implications for the ESS}
\label{tab:ess_results}
\begin{tabular}{cccccccc} \hline\hline
 & \multicolumn{2}{c}{$Y_R$ ($M_\odot$)} & & & \multicolumn{3}{c}{$(N_{R}/N_{I})_{\rm ESS}$} \\  
 $R/I$ & This Work & Banerjee et al. 2016 & $\tau_R$ (Myr) & $X_{I,\odot}$ & Data & Uniform Injection & Tiered Injection\\ \hline
$^{10}$Be/$^{9}$Be & $6.08 \times 10^{-11} $ & $3.26 \times 10^{-10} $ & $2.00 $ & $1.40 \times 10^{-10} $ & $(7.5\pm 2.5) \times 10^{-4} $ & $1.00 \times 10^{-6} $ & $4.50 \times 10^{-5} $ \\
$^{26}$Al/$^{27}$Al & $6.18 \times 10^{-6} $ & $2.91 \times 10^{-6} $ & $1.03 $ & $5.65 \times 10^{-5} $ & $(5.23\pm 0.13) \times 10^{-5} $ & $2.13 \times 10^{-7} $ & $8.73 \times 10^{-6} $\\
$^{36}$Cl/$^{35}$Cl & $3.26 \times 10^{-6} $ & $1.44 \times 10^{-7} $ & $0.434 $ & $3.50 \times 10^{-6} $ & $(2.44\pm 0.65)\times 10^{-5} $ & $6.88 \times 10^{-7} $ & $1.20 \times 10^{-6} $ \\
$^{41}$Ca/$^{40}$Ca & $7.11 \times 10^{-6} $ & $3.66 \times 10^{-7} $ & $0.147 $ & $5.88 \times 10^{-5} $ & $(4.6\pm 1.9) \times 10^{-9} $ & $4.31 \times 10^{-9} $ & $5.24 \times 10^{-9} $ \\
$^{53}$Mn/$^{55}$Mn & $2.73 \times 10^{-5} $ & $1.22 \times 10^{-5} $ & $5.40 $ & $1.29 \times 10^{-5} $ & $(7\pm 1) \times 10^{-6} $ & $6.97 \times 10^{-6} $ & $6.45 \times 10^{-6} $ \\
$^{60}$Fe/$^{56}$Fe & $4.17 \times 10^{-6} $ & $3.08 \times 10^{-6} $ & $3.78 $ & $1.12 \times 10^{-3} $ & $(1.01\pm0.27) \times 10^{-8}$ & $1.04 \times 10^{-8} $ & $1.05 \times 10^{-8} $ \\
$^{107}$Pd/$^{108}$Pd & $1.42 \times 10^{-10} $ & $1.37 \times 10^{-10} $ & $9.38 $ & $9.92 \times 10^{-10} $ & $(6.6\pm 0.4) \times 10^{-5} $ & $4.85 \times 10^{-7} $ & $ 6.43 \times 10^{-5} $ \\
$^{135}$Cs/$^{133}$Cs & $3.06 \times 10^{-10} $ & $2.56 \times 10^{-10} $ & $3.32 $ & $1.24 \times 10^{-9} $ & $\sim5\times 10^{-4} $, $< 2.8 \times 10^{-6} $ & $7.12 \times 10^{-7} $ & $1.40 \times 10^{-5}$ \\
$^{182}$Hf/$^{180}$Hf & $9.46 \times 10^{-12} $ & $8.84 \times 10^{-12} $ & $12.84 $ & $2.52 \times 10^{-10} $ & $(1.018\pm 0.043) \times 10^{-4} $ & $1.27 \times 10^{-7} $ & $2.48 \times 10^{-7} $ \\
$^{205}$Pb/$^{204}$Pb & $1.01 \times 10^{-10} $ & $9.20 \times 10^{-11} $ & $24.96 $ & $3.47 \times 10^{-10} $ & $(1.8\pm 1.2) \times 10^{-3}$ & $1.02 \times 10^{-6} $ & $6.20 \times 10^{-5} $ \\ \hline
\end{tabular} 
\tablecomments{The data references are $^{10}$Be \citep{2000Sci...289.1334M,2003GeCoA..67.3165M,2012ApJ...748L..25W,2013E&PSL.374...11S}, 
    $^{26}$Al \citep{1976GeoRL...3..109L,2008E&PSL.272..353J},
    $^{36}$Cl \citep{Tang.Liu.ea:2017,Lugaro.Ott.ea:2018},
    $^{41}$Ca \citep{Liu:2017},
    $^{53}$Mn \citep{Tranquier.Birck.ea:2008,Tissot.Dauphas.ea:2017},
    $^{60}$Fe \citep{Tang.Dauphas:2015},
    $^{107}$Pd \citep{Matthes.Goedde.ea:2018}
    $^{135}$Cs \citep{2001E&PSL.193..459H,Brennecka.Kleine.ea:2017},
    $^{182}$Hf \citep{2008GeCoA..72.6177B,Kruijer.Kleine.ea:2014} and
    $^{205}$Pb \citep{Palk.Andreasen.ea:2018}. 
Our yield of $^{10}$Be has large uncertainties (see \S\ref{sec:be10}).
The yields of $^{182}$Hf assume the stellar decay rate of $^{181}$Hf. 
Using the laboratory rate, \citet{Banerjee.Qian.ea:2016} obtained a higher yield of 
$4.04\times10^{-11}\,M_\odot$.}
\end{table*}

\begin{figure}
    \centering
    \includegraphics[width=\linewidth]{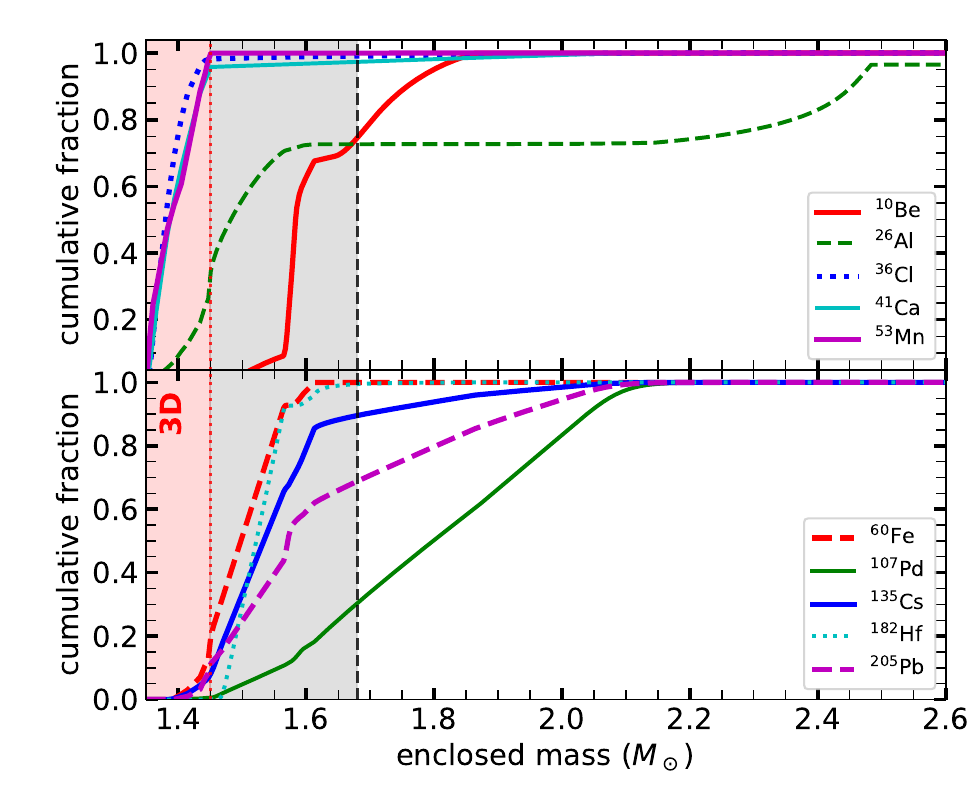}
    \caption{Cumulative fraction of the yield for each SLR as a function
of the enclosed mass for our combined model. The dotted line at $M_{\rm tr}=1.45\,M_\odot$
indicates the transition between the inner and outer ejecta. In a hypothetical scenario,
the dashed line at $M_{\rm mix}=1.68\,M_\odot$ represents the boundary inside which only a 
small fraction of the ejecta might have been injected into the proto-solar cloud.}
    \label{fig:cumulative_fraction}
\end{figure}

In this section we discuss the yields of SLRs from the $11.8\,M_\odot$ CCSN based on
our combined model and the implications for the abundances of these SLRs in the ESS.
We give these yields in Table~\ref{tab:ess_results} along with those obtained by
\citet{Banerjee.Qian.ea:2016} using a 1D model.\footnote{ 
Our results are consistent with the ESS data on $^{92}$Nb, $^{97}$Tc, $^{98}$Tc, 
$^{126}$Sn, $^{129}$I, $^{146}$Sm, $^{244}$Pu, and $^{247}$Cm 
\citep[e.g.,][]{Lugaro.Ott.ea:2018} in that these SLRs would not be overproduced. 
In Table~\ref{tab:ess_results}, we focus on those SLRs discussed by 
\citet{Banerjee.Qian.ea:2016} for comparison with that study.}
It can be seen that our yields and
theirs are close for $^{60}$Fe, $^{107}$Pd, $^{135}$Cs, $^{182}$Hf, and $^{205}$Pb,
but differ greatly for $^{10}$Be, $^{26}$Al, $^{36}$Cl, $^{41}$Ca, and $^{53}$Mn.
The difference for $^{10}$Be is discussed in detail in \S\ref{sec:be10}. 
The similarities and differences for the other SLRs can be traced to their regions 
of production. Figure~\ref{fig:cumulative_fraction} shows the cumulative fraction
of the yield for each SLR as a function of the enclosed mass for our combined model.
Because the transition between the inner and outer ejecta is at the enclosed mass of
$M_{\rm tr}=1.45\,M_\odot$, almost all of the $^{36}$Cl, $^{41}$Ca, and $^{53}$Mn
are attributed to the inner ejecta modeled by our 3D simulation. In contrast, nearly 
all of the $^{60}$Fe, $^{107}$Pd, $^{135}$Cs, $^{182}$Hf, and $^{205}$Pb are attributed 
to the outer ejecta covered by our adopted 1D model. Whereas the details of the explosion, 
especially the mass cut, have a large impact on the production of the SLRs predominantly 
attributed to the inner ejecta, the yields of those SLRs predominantly attributed to the 
outer ejecta largely reflect the weak s-process nucleosynthesis during the pre-CCSN 
evolution and are not significantly affected by the explosion. The mass cut determined 
from our 3D simulation is at the enclosed mass of $M_{\rm cut}=1.35\,M_\odot$. 
\citet{Banerjee.Qian.ea:2016}, however, chose a significantly larger mass cut at 
$M_{\rm cut}'\approx 1.44\,M_\odot$. Therefore, their 1D model missed the majority of 
the SLR yields attributed to our inner ejecta while giving very similar results for 
those attributed to our outer ejecta. In addition, the difference in the mass cut
accounts for that in the $^{26}$Al yield because $\approx 35\%$ of this yield is 
attributed to our inner ejecta. 

As discussed in \S\ref{sec:be10}, our nominal $^{10}$Be yield has large uncertainties
due to the uncertain rate of $^{10}$Be$(p,\alpha)^{7}$Li and possible effects of
neutrino flavor oscillations (see Table~\ref{tab:be10_sensitivity}). For completeness,
we also explore the latter effects on the yields of the other SLRs. 
In the absence of flavor oscillations, we find that the yields of all SLRs from our 
adopted 1D model are close to those from our combined model.
To estimate the effects of flavor oscillations, we use the 1D model to consider
two extreme cases of either $\bar{\nu}_\mathrm{e}\leftrightarrow\bar{\nu}_x$ or 
$\nu_\mathrm{e}\leftrightarrow\nu_x$ oscillations as done for $^{10}$Be. In either case,
there is very little change to the yields of $^{53}$Mn, $^{60}$Fe, $^{107}$Pd, 
$^{135}$Cs, $^{182}$Hf, and $^{205}$Pb. For $\bar{\nu}_\mathrm{e}\leftrightarrow\bar{\nu}_x$
oscillations, significant changes are a factor of $\sim 2.3$ increase in the
$^{36}$Cl yield due to the enhancement of $^{36}$Ar$(\bar\nu_\mathrm{e},\mathrm{e}^+)^{36}$Cl
and a factor of $\sim 1.4$ increase in the $^{41}$Ca yield due to the enhancement 
of $^{41}$K$(\bar\nu_\mathrm{e},\mathrm{e}^+)^{41}$Ca. For $\nu_\mathrm{e}\leftrightarrow\nu_x$ oscillations,
the only significant change is a factor of $\sim 1.9$ increase in the $^{26}$Al yield 
due to the enhancement of $^{26}$Mg$(\nu_\mathrm{e},\mathrm{e}^-)^{26}$Al.

Based on their yields, \citet{Banerjee.Qian.ea:2016} found that an $11.8\,M_\odot$ CCSN
could account for the abundances of $^{10}$Be, $^{41}$Ca, and $^{107}$Pd in the ESS.
For agreement with the data on $^{53}$Mn, they proposed that only 1.5\% of the innermost 
$\sim 0.01$--$0.12\,M_\odot$ of the ejecta in their model might have
been ejected due to fallback. In this scenario, however, the CCSN contribution
to $^{60}$Fe is only consistent with the values of 
$(^{60}$Fe/$^{56}$Fe$)_{\rm ESS}\sim (5$--$10)\times 10^{-7}$ reported by
\citet{Mishra.Goswami:2014}, which greatly exceed those of
$(^{60}$Fe/$^{56}$Fe$)_{\rm ESS}=(1.01\pm0.27)\times 10^{-8}$ reported by
\citet{Tang.Dauphas:2015}. Because neither our 3D simulation nor our 1D model shows any 
indication of fallback and our nucleosynthesis calculations give significantly different yields 
of $^{10}$Be, $^{26}$Al, $^{36}$Cl, $^{41}$Ca, and $^{53}$Mn (see Table~\ref{tab:ess_results}), 
below we reexamine the possible contributions from the $11.8\,M_\odot$ CCSN to the SLRs in the 
ESS. We focus on our yields in Table~\ref{tab:ess_results} without considering neutrino flavor 
oscillations. Such considerations do not change our essential results, but are commented on.

\subsection{Implications for the ESS: Uniform Injection}
Assuming that only the $11.8\,M_\odot$ CCSN contributed a radioactive isotope $R$ 
with mass number $A_R$ and lifetime $\tau_R$ and that
long-term production by other sources provided its stable reference isotope 
$I$ with mass number $A_I$ to 
the proto-solar cloud, we calculate the number ratio of $R$ to $I$ in the ESS as 
\begin{equation}
    \label{eq:ess_ratio}
    \left( \frac{N_R}{N_I} \right)_{\mathrm{ESS}} 
    =\frac{fY_R/A_R}{X_{I,\odot}M_\odot/A_I}\mathrm{exp}\left( -\Delta/\tau_R \right),
\end{equation}
where $Y_R$ is the CCSN yield of $R$, $X_{I,\odot}$ is the solar mass fraction of $I$,
$f$ is the fraction of the $R$ yield injected into the proto-solar cloud,
and $\Delta$ is the interval between the production of $R$ by the CCSN and its 
incorporation into ESS solids.

\begin{figure}
    \centering
    \includegraphics[width=\linewidth]{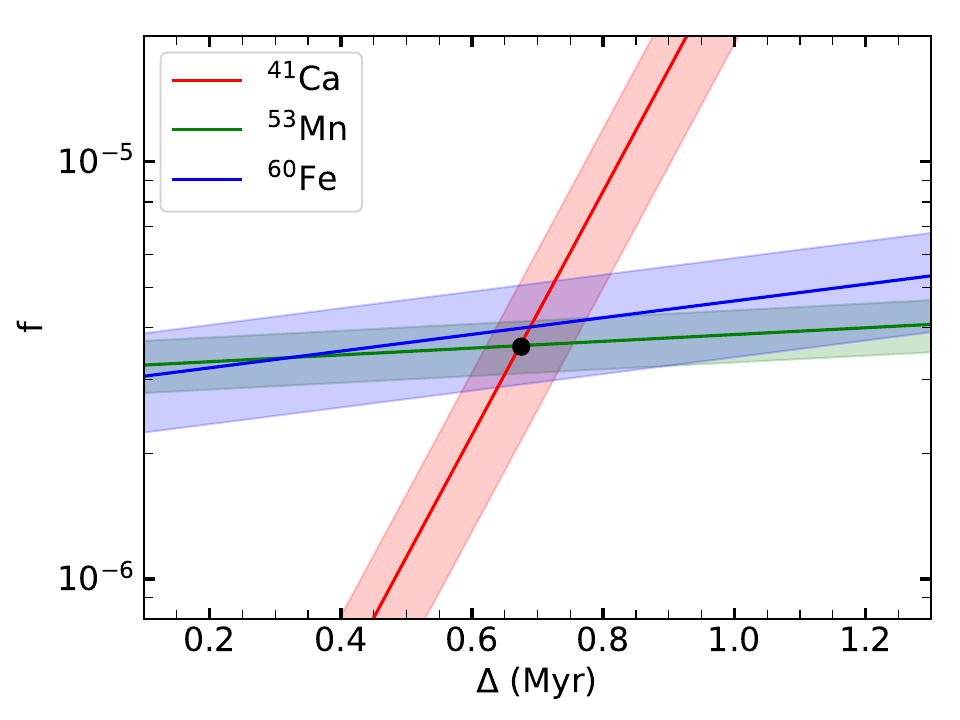}
    \caption{Determination of $f$ and $\Delta$ for the uniform injection scenario.
    The band for an SLR shows combinations of $f$ and $\Delta$ that allow the result from 
    Equation~(\ref{eq:ess_ratio}) to account for the corresponding data. The black dot 
    indicates the ``best-fit'' values of $f=3.6\times 10^{-6}$ and $\Delta=0.675$~Myr, for
    which simultaneous agreement with the data on $^{41}$Ca, $^{53}$Mn, and $^{60}$Fe can be
    obtained.}
    \label{fig:f_delta}
\end{figure}

It is reasonable to assume that $\Delta$ is the same for all of the SLRs.
We first consider the uniform injection scenario where they also have the same $f$.
We search for $f$ and $\Delta$ with which the results from Equation~(\ref{eq:ess_ratio})
can account for the data on some of the SLRs without exceeding those on the rest.
As shown in Figure~\ref{fig:f_delta} and Table~\ref{tab:ess_results}, for the
``best-fit'' values of $f=3.6\times 10^{-6}$ and $\Delta=0.675$~Myr, the
$11.8\,M_\odot$ CCSN can account for the data on $^{41}$Ca, $^{53}$Mn, and $^{60}$Fe
while making negligible contributions to the other SLRs.\footnote{
Even the very low upper limit ($^{135}$Cs/$^{133}$Cs)$_{\rm ESS}<2.8\times10^{-6}$ 
from \citet{Brennecka.Kleine.ea:2017} can be satisfied.} Note that we have adopted the 
low values of $(^{60}$Fe/$^{56}$Fe$)_{\rm ESS}=(1.01\pm0.27)\times 10^{-8}$ reported by
\citet{Tang.Dauphas:2015}, which appear to be supported a the recent study of
\citet{Trappitsch.Boehnke.ea:2018}. Note also that simultaneous agreement with the 
data on $^{53}$Mn and $^{60}$Fe can be obtained for a relatively narrow range of $f$ 
but a wide range of $\Delta$. Effectively, this agreement determines the best-fit 
value of $f$ while the additional agreement with the data on the very short-lived 
$^{41}$Ca determines the best-fit value of $\Delta$. Because neutrino flavor oscillations
have little impact on the yields of $^{53}$Mn and $^{60}$Fe and only a modest effect 
(a factor of $\sim 1.4$) on the $^{41}$Ca yield, they do not affect the above results
significantly.

\subsection{Implications for the ESS: Tiered Injection}
\label{sec:mixing_scenario}
The uniform injection scenario is commonly assumed in discussing the contributions from
a CCSN to the SLRs in the ESS. This scenario appears to be supported by the simulations of 
\citet{Foster.Boss.ea:1998}, who found that material lagging far behind but moving with a 
similar velocity to the shock can be injected into a proto-stellar cloud with a similar 
efficiency to the material in the shock front. This result, however, applies only 
to the lagging material that can move to the shock-cloud interface when injection is still
occurring through the Rayleigh-Taylor-like clumps. Simulations by \citet{1998ApJ...500...95T}
showed that in addition to a forward shock, there is also a reverse shock during the 
evolution of a CCSN remnant. This feature leads to a complicated velocity profile, sometimes 
including negative velocities, for the material behind the forward shock. In view of such
complications, we also consider a tiered injection scenario with a hypothetical boundary 
at the enclosed mass of $M_{\rm mix}$ for the CCSN ejecta. In this scenario, only a small 
fraction $\epsilon$ of the ejecta inside the boundary is mixed with the outside ejecta
and injected into the proto-solar cloud with the same efficiency. Effectively, a fraction
$f$ of the yield in the outside ejecta is injected for each SLR, while the injected fraction for
the inside ejecta is $\epsilon f$.

As an example, we choose $M_{\mathrm{mix}}=1.68\,M_\odot$, $\epsilon=0.5\,\%$,
$f=7\times 10^{-4}$, and $\Delta=0.9\,\mathrm{Myr}$.
With these parameters, we can match the data on $^{41}$Ca, $^{53}$Mn, $^{60}$Fe, and
$^{107}$Pd without exceeding those on the other SLRs\footnote{
The possible exception is $^{135}$Cs, for which our result is compatible with
($^{135}$Cs/$^{133}$Cs)$_{\rm ESS}\sim 5\times10^{-4}$ from \citet{2001E&PSL.193..459H}
but in conflict with ($^{135}$Cs/$^{133}$Cs)$_{\rm ESS}<2.8\times10^{-6}$ 
from \citet{Brennecka.Kleine.ea:2017}. The latter, however, was inferred from the lack of 
$^{135}$Ba deficits in volatile-depleted samples, in contrast to the former, which was
based on measurements of the $^{135}$Ba excess from the $^{135}$Cs decay in different samples. 
More studies are required to resolve the above discrepancy.} (see Table~\ref{tab:ess_results}).
This result is not affected by neutrino flavor oscillations. We have also checked that
the large uncertainties in the $^{10}$Be yield would allow the highest possible value
of $(^{10}$Be/$^9$Be$)_{\rm ESS}$ to be $\approx 2\times 10^{-4}$, which is still
a factor of $\approx 2.5$ below the lower end of the observed values. Overall,
the above result is close to what \citet{Banerjee.Qian.ea:2016} tried to achieve
with similar values of $f$ and $\Delta$ but by invoking fallback that we do not
find in our 3D simulation or 1D model. Note also that our $^{41}$Ca yield is $\approx 19$
times as high as theirs and we have adopted the much lower values of 
\citet{Tang.Dauphas:2015} for the data on $^{60}$Fe. 

\section{Discussion and Conclusions}
\label{sec:conclusions}
We have studied the nucleosynthesis of an $11.8\,M_\odot$ CCSN based on
a 3D simulation of the inner ejecta. Accounting for the uncertainties in the neutrino transport 
with reasonable corrections to the neutrino emission properties, we have found mostly proton-rich
inner ejecta that provide substantial yields of $^{45}$Sc and $^{64}$Zn. 
A proper assessment of the production of heavier isotopes, however,
requires improvement of our simulation by incorporating
more accurate neutrino transport to determine self-consistently the $Y_\mathrm{e}$ of the 
neutrino-heated ejecta. Combining the results for the inner ejecta with those for the
outer ejecta from a suitable 1D model, we illustrate a method to obtain complete CCSN yields 
from a 3D simulation that only covers a limited central region. Such yields are important for 
chemical evolution considerations and for comparison to observational signatures such as the 
abundances of SLRs in the ESS.

Using our yields of the SLRs, we have explored the possibility that an $11.8\,M_\odot$ CCSN
might have triggered the formation of the solar system and provided some of the SLRs in the
ESS. In particular, we have discussed the uniform injection scenario, which can account for 
the data on $^{41}$Ca, $^{53}$Mn, and $^{60}$Fe without exceeding those on the other SLRs,
and the tiered injection scenario, which can account for the data on one more SLR, $^{107}$Pd.
The latter scenario is close to what \citet{Banerjee.Qian.ea:2016} tried to achieve. 
Our 3D simulation and 1D model, however, show no indication of the fallback that they invoked.
Compared to their study, we have also identified large uncertainties in the production of $^{10}$Be,
obtained significantly larger yields of $^{26}$Al, $^{36}$Cl, $^{41}$Ca, and $^{53}$Mn, and
adopted the much lower values of \citet{Tang.Dauphas:2015} for the data on $^{60}$Fe, 
which appear to be supported by the more recent study of \citet{Trappitsch.Boehnke.ea:2018}. 

Finally, we emphasize that whereas our tiered injection scenario superficially looks like 
the usual scenario of mixing and fallback of the innermost CCSN matter 
\citep[e.g.,][]{Takigawa.Miki.ea:2008,Banerjee.Qian.ea:2016}, the underlying physics
and processes are very different. The usual scenario is closely related to 
the explosion mechanism that determines the mass cut and hence the actual amount
of the ejecta, which are then injected into the proto-solar cloud with the same efficiency
as in our uniform injection scenario. In contrast, our tiered injection scenario is 
motivated by the velocity profile during the CCSN remnant evolution, which could 
allow the inner part of the ejecta to be injected into the proto-solar cloud with
a much lower efficiency than the outer part. Of course, the applicable scenario
must ultimately be determined by detailed simulations of the CCSN and its subsequent
interaction with the ISM including the proto-solar cloud. Combining results
from observations and hydrodynamic simulations, \citet{Krause.Burkert.ea:2018} recently 
showed that CCSNe play a major role in shaping the ISM and influencing subsequent star 
formation in Scorpius--Centaurus OB2. It would be interesting to study if
our low-mass CCSN with a lifetime of $\sim 19$~Myr could have triggered the collapse of
the proto-solar cloud in a similar fashion and if earlier CCSNe in the associated
star formation region could have provided some SLRs such as $^{26}$Al. Analyses like those 
presented here need to be supported by such studies to test any candidate CCSN model
as a trigger for the formation of the solar system.

\begin{acknowledgments}
We thank the anonymous referee for constructive and helpful comments. 
This work was supported in part by the US Department of Energy [DE-FG02-87ER40328 (UM)]. 
B.M. was supported by ARC Future Fellowship FT160100035. This research
was undertaken with the assistance of resources and services from the
National Computational Infrastructure, which is supported by the
Australian Government.  It was supported by resources provided by the
Pawsey Supercomputing Centre with funding from the Australian
Government and the Government of Western Australia.
Further calculations were run at the GSI Helmholtz Centre for Heavy Ion Research and the Minnesota 
Supercomputing Institute.
\end{acknowledgments}

\software{CoCoNuT-FMT \citep{Mueller.Tauris.ea:2019,Mueller.Janka.ea:2015},
KEPLER \citep{Weaver.Zimmerman.Woosley:1978}, Matplotlib \citep{Hunter:2007}.}

\bibliographystyle{apjnew}

\end{document}